\documentclass[twocolumn,trackchanges]{aastex63}

\DeclareRobustCommand{\ion}[2]{\textup{#1\,\textsc{\lowercase{#2}}}}

\received{November 16, 2020}
\revised{--}
\accepted{--}

\submitjournal{ApJS}

\shorttitle{A Scaling Between SP and HMI to Improve Field Extrapolations}
\shortauthors{Beck et al.}

\begin{document}

\title{Derivation and Application of a Scaling Between Hinode/SP and SDO/HMI Vector Magnetic Fields to Improve Magnetic Field Extrapolations}

\correspondingauthor{CB}

\author[0000-0001-7706-4158]{C. Beck}
\affiliation{National Solar Observatory (NSO), \\
3665 Discovery Drive, Boulder, CO 80303, USA}

\author[0000-0003-0819-464X]{A. Prasad}
\affiliation{Rosseland Centre for Solar Physics, University of Oslo. \\
Postboks 1029, Blindern NO-0315, Oslo, Norway}
\affiliation{Institute of Theoretical Astrophysics, University of Oslo, \\
Postboks 1029, Blindern NO-0315, Oslo, Norway}

\author[0000-0002-7570-2301]{Q. Hu}
\affiliation{Department of Space Science,
The University of Alabama in Huntsville (UAH), \\
Huntsville, AL 35805, USA}
\affiliation{Center for Space Plasma and Aeronomic Research (CSPAR),
The University of Alabama in Huntsville (UAH), \\
Huntsville, AL 35805, USA}

\author[0000-0002-8496-0353]{M. S. Yalim}
\affiliation{Center for Space Plasma and Aeronomic Research (CSPAR),
The University of Alabama in Huntsville (UAH), \\
Huntsville, AL 35805, USA}

\author[0000-0002-5504-6773]{S. Gosain}
\affiliation{National Solar Observatory (NSO), \\
3665 Discovery Drive, Boulder, CO 80303, USA}

\author[0000-0002-9308-3639]{D. Prasad Choudhary}
\affiliation{Department of Physics and Astronomy, \\
California State University, Northridge (CSUN), CA 91330-8268, USA}

\begin{abstract}
Full-disk measurements of the solar magnetic field by the Helioseismic and Magnetic Imager (HMI) are often used for magnetic field extrapolations, but its limited spatial and spectral resolution can lead to significant errors. We compare HMI data with observations of NOAA 12104 by the Hinode Spectropolarimeter (SP) to derive a scaling curve for the magnetic field strength, $B$. The SP data in the \ion{Fe}{i} lines at 630\,nm were inverted with the SIR code. We find that the Milne-Eddington inversion of HMI underestimates $B$ and the line-of-sight flux, $\Phi$, in all granulation surroundings by an average factor of 4.5 in plage and 9.2 in the quiet Sun in comparison to the SP. The deviation is inversely proportional to the magnetic fill factor, $f$, in the SP results. We derived a correction curve to match the HMI $B$ with the effective flux $B\,f$ in the SP data that scaled HMI $B$ up by 1.3 on average. A comparison of non-force-free field extrapolations over a larger field of view without and with the correction revealed minor changes in connectivity and a proportional scaling of electric currents and Lorentz force ($\propto B \sim 1.3$) and free energy ($\propto B^2\sim 2$). Magnetic field extrapolations of HMI vector data with large areas of plage and quiet Sun will underestimate the photospheric magnetic field strength by a factor of 5--10 and the coronal magnetic flux by at least 2. An HMI inversion including a fill factor would mitigate the problem.  
\end{abstract}


\section{Introduction}
Magnetic fields on the Sun are primarily observed by measuring the Zeeman splitting of spectral lines in a magnetized medium \citep{hale1908}. The Stokes profiles of spectral lines contain the information about thermodynamic (temperature, density, velocity) and magnetic properties (magnetic field vector) and their gradients along the line of sight (LOS) in the intensity spectra and the Zeeman-split polarization components \citep[e.g.,][]{deltoroiniesta+ruizcobo2016}. In the past decades, several attempts with a varying degree of sophistication have been made to derive these physical properties of the solar atmosphere from observed Stokes profiles by applying inversion techniques \citep{cobo+toroiniesta1992,asensio+etal2008,borrero+etal2011, socas-navarro+etal2015,beck+etal2019a,delacruzrodriguez+etal2019,sainzdalda+etal2019,ruizcobo+etal2022}. It was found that the magnetic vector fields obtained with different spectral lines and different spatial and spectral resolution are not the same, especially if the small-scale magnetic fields on the solar surface ($< 1^{\prime\prime}$) are spatially unresolved or the thermal broadening (3-6\,pm) or the Zeeman splitting is spectrally unresolved \citep[e.g.,][]{berger+lites2003,pietarila+etal2013,sainzdalda2017,wang+etal2022}. 

\begin{figure}
\resizebox{8.8cm}{!}{\includegraphics{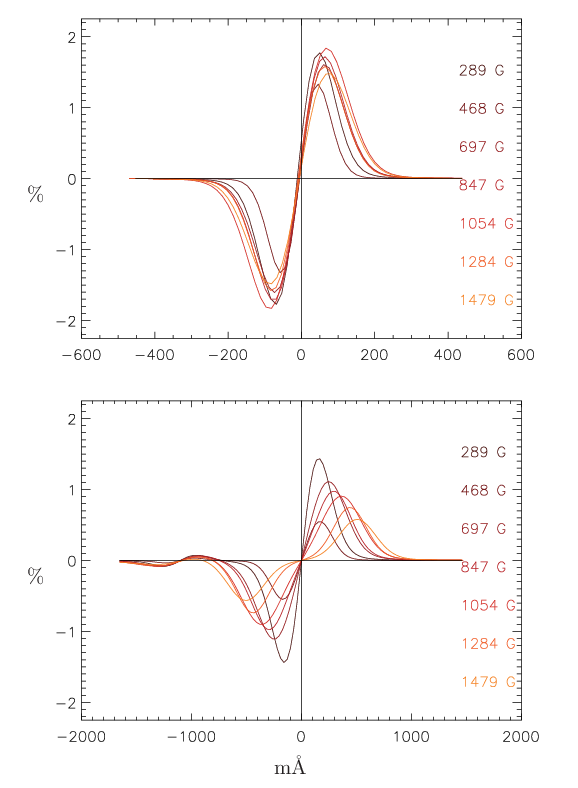}}
\caption{Synthetic \ion{Fe}{i} 1564.8 nm (bottom panel) and \ion{Fe}{i} 630.25 nm (top panel) Stokes $V$ spectra for different field strengths (given at the right-hand side) and a varying magnetic fill factor.}\label{fig0}
\end{figure}

The magnetic field strengths derived from either space-based or ground-based observations in different Fraunhofer lines are not identical as the lines generally form at different heights in the stratified solar atmosphere \citep{grossmanndoerth1994,wenzler+etal2004,cabrera+bellot+iniesta2005}. The underlying transitions also differ in their response to the magnetic field strength due to their specific Land{\'e} coefficients \citep[e.g.,][]{harvey1973} and their rest wavelengths, where for spectral lines in the weak-field limit primarily the polarization amplitudes change instead of the wavelength separation of polarization lobes \citep[e.g.,][]{jefferies+etal1989}. Figure \ref{fig0} shows these effects in synthetic Stokes $V$ spectra of \ion{Fe}{i} 1564.8 nm and \ion{Fe}{i} 630.25 nm. The magnetic field strengths and fill factors \citep[e.g.,][]{beck+rezaei2009} used here resulted from an inversion of simultaneous observations of both lines in the quiet Sun \citep{martinez+etal2008}, where the 630.25\,nm spectra were selected to have about the same polarization amplitude of 1.8\,\% in the Stokes $V$ lobes. While the spectra at 1564.8\,nm show a clear variation in shape, splitting and amplitude, the 630.25\,nm spectra stay similar to undistinghuishable. For magnetographs, this behavior can lead to significantly different results depending on which line is observed at which spectral resolution, or in the other direction, give the same result for different true magnetic properties because of the similarity of the spectra in the visible (top panel of Figure \ref{fig0}), especially in the presence of noise \citep{deltoroiniesta+etal2010,deltoroiniesta+etal2012}.

Magnetographs and instruments with a lower spatial or spectral resolution usually give systematically lower field strengths than those with a higher resolution because of the effect of the magnetic fill factor inside the resolution element for unresolved magnetic fields \citep{plowman+berger2020}. \citet{fouhey+etal2023} found that even mismatches in spatial scaling can lead to different results from comparing measurements by the Helioseismic and Magnetic Imager \citep[HMI;][]{scherrer+etal2012} and the Hinode/Solar Optical Telescope Spectropolarimeter \citep[SP;][]{kosugi+etal2007,ichimoto+etal2008,tsuneta+etal2008}.

The determination of the magnetic field strength of different solar structures in the photosphere and at other heights in the solar atmosphere is a focus of contemporary solar physics due to its various applications. For example, the derivation of the magnetic field at different heights by various magnetic extrapolation schemes uses primarily photospheric magnetograms as bottom boundary conditions \citep{wiegelmann2008,miwawaki+etal2016,wiegelmann+sakurai2021,vissers+etal2022}. Extrapolated magnetic fields are the primary ingredients to estimate the energy contained in flares or coronal mass ejections \citep[][and references therein]{yalim+etal2020,yu+etal2023}. The magnetic field strength values at the boundaries will then naturally change the vertical structure and the energy content that is derived with these extrapolation techniques. 

\citet{marchenko+etal2022} found that variations in the total solar irradiance of the Sun at times of solar activity minima seem to follow the trend in total magnetic flux from sources with $|B| > 80$\,G, which couples a global solar property to photospheric magnetic fields. The derivation of electric currents \citep{puschmann+etal2010a,borrero+etal2023} in the solar atmosphere above the photosphere that are relevant for heating processes \citep{louis+etal2021,santos+etal2022,yalim+etal2023,yalim+etal2024} is often based on magnetic field extrapolations. Hence, a reliable determination of field strengths or methods for consoling magnetograms obtained using different instruments with each other are very useful. 

The LOS magnetic flux inferred from calibrated Michelson Doppler Interferometer \citep[MDI;][]{scherrer+etal1995} data was found to be larger than the one derived from HMI data by a factor of 1.40 with an additional change depending on the heliocentric angle \citep{liu+etal2012}. \citet{riley+etal2014} determined similar scaling factors for the LOS magnetic flux from a comparison of different ground-based and space-based observations with different spatial and spectral resolution. \citet{sun+etal2022} used deep-learning models to improve the stability of HMI magnetograms by coupling them to images from the Atmospheric Imaging Assembly \citep[AIA;][]{lemen+etal2012}. \citet{virtanen+mursula2017} used a scaling of the coefficients of a harmonic expansion of the magnetic field to match different data sources, where only the first few terms were relevant for coronal modeling. Opposite to other scaling methods, their approach can be easily applied to data sets of different spatial resolution. 

In a more direct pixel-to-pixel comparison of Hinode SP and MDI data \citet{moon+etal2007} found that the MDI magnetic flux density could be underestimated by a factor of about two with additional deviations from the SP in the umbra because of Zeeman saturation. \citet{kontogiannis+etal2011} found deviations by up to a factor of five between the same instruments in quiet Sun (QS) regions. Instead of determining a relative scaling, \citet{higgins+etal2022} combined data from HMI and SP using multiple convolutional neural networks to derive photospheric magnetic fields in a unified inversion scheme based on both data sources. The introduction of a magnetic fill factor into the HMI inversion produced significant differences in derived magnetic properties in plage regions, where spatially unresolved -- at both HMI and SP resolution -- magnetic elements are expected. 

The purpose of the current investigation is twofold. On the one hand, we want to determine a scaling curve to improve the magnetic field strength derived from HMI observations from a comparison to simultaneous observations with the Hinode SP that can afterwards be applied to any HMI data set, and on the other hand we want to estimate the differences in the magnetic field  properties at different heights derived using scaled and non-scaled photospheric magnetograms in a subsequent magnetic field extrapolation, similar to the effort in \citet{kontogiannis+etal2011}. The primary motivation for the latter is a possible application to attribute formation heights to magnetic field strength values derived from an inversion of the chromospheric \ion{He}{i} line at 1083\,nm, while the scaling curve could potentially be of benefit to any study based on HMI field strengths. In difference to \citet{sainzdalda2017} we do not want to trace in detail where the eventual differences between HMI and the SP arise from but primarily aim for a possible improvement of the standard HMI vector field data product.
  
Section \ref{sec_obs} describes the data sets used. Section \ref{sec_ana} explains the analysis methods employed. Our results are given in Section \ref{sec_res} and summarized in Section \ref{sec_summ}. We discuss the findings in Section \ref{sec_disc}, while Section \ref{sec_conc} provides our conclusions.
   
\section{Observations  \label{sec_obs}}
\subsection{Hinode SP Data}
We used observations of the active region (AR) NOAA 12104 on 2014 July 3 when it was at a heliocentric angle of about 16 degrees. The Hinode SP scanned both polarities of the AR from UT 19:11--19:46 on 560 steps of 0\farcs3 step width for a total field of view (FOV) of about $168^{\prime\prime} \times 115^{\prime\prime}$. The integration time was 3.8\,s per step. The spectral window covered a range from 630.0 to 630.3\,nm with a spectral sampling of 2.15 pm\,pix$^{-1}$, while the spatial sampling along the slit was about 0\farcs3\,pix$^{-1}$. 
\subsection{HMI Data}
To obtain co-aligned SP and HMI data, we used HMI full-disk intensity observations at 45\,s cadence and vector magnetic field data at 12\,min cadence taken between UT 19 and UT 20 and cut out the SP FOV. The HMI spatial sampling is about 0\farcs5\,pix$^{-1}$ with a spectral sampling of 7 pm on 6 different wavelength positions for the full vector mode \citep{schou+etal2012}. These data were used to derive the scaling curve between HMI and SP.

For the application of the HMI scaling curve, we downloaded a Space-Weather HMI Active Region Patches \citep[SHARP;][]{bobra+etal2014} cut-out of the HMI vector magnetic field data at UT 19:24 of a larger $665^{\prime\prime} \times 426^{\prime\prime}$ FOV that included most to all of NOAA 12104 and NOAA 12107 to the South-East of the former. These HMI data were cylindrical equal area (CEA) projected to square pixels on the solar surface. The SP FOV is fully included in this HMI cut-out.

We also obtained the same HMI SHARP CEA data after a correction for scattered light.  This correction was done by the HMI team and employs a deconvolution with a point spread function (PSF). This PSF has the form of a Lorenztian convolved with an Airy function and was determined from pre-launch calibration observations and post-launch Venus transit and lunar eclipse data. The deconvolution uses a Richardson-Lucy algorithm and takes less than one second per full-disk image (Norton et al, in prep). In March 2018, the HMI team began providing these data to the public on a daily basis. 

\begin{figure*}
\centerline{\resizebox{17.cm}{!}{\includegraphics{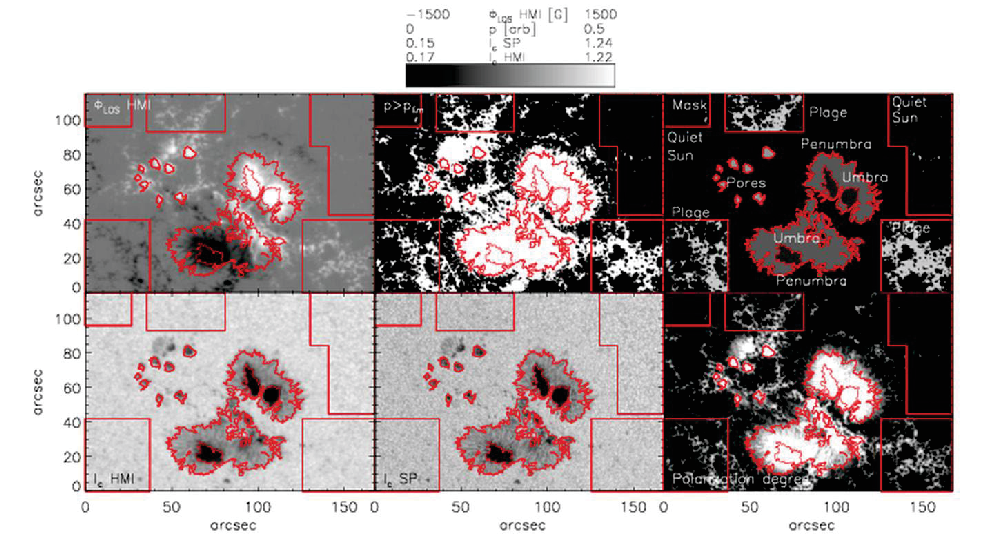}}}
\caption{Overview of NOAA 12104 on 2014 Jul 03 at about UT 19:30. Bottom row, left to right: continuum intensity $I_c$ from HMI pseudo-scan, $I_c$ and polarization degree $p$ from SP data. Top row, left to right: LOS magnetic flux $\Phi_{LOS}$ from HMI, locations with $p>p_{lim}$ in SP data, and mask of surface structures (QS to umbra from white to dark grey). Red rectangles at the corners indicate QS and plage regions, other contour lines indicate pores, penumbral and umbral regions.}\label{fig1}
\end{figure*}   
\subsection{Spatial Alignment}
To co-align the SP and HMI data, we constructed a "pseudo-scan" of HMI data that mimics stepping a virtual slit across a static two-dimensional HMI FOV \citep[][their Appendix B.~2]{beck+etal2007}. We first forced the HMI intensity images at 45\,s cadence from UT 19--20 to a common fixed FOV centered on NOAA 12104, which corresponds to the conditions during the SP data acquistion with active tip-tilt image stabilization. We set the assumed start positions of the SP slit $x_0$ and $y_0$ inside the HMI FOV and increased its position in $x$ by 0\farcs291 with each step. For each SP step, a slice with the length of the SP slit in $y$ was then cut out from the HMI image closest in time. We adjusted the initial positions $x_0$ and $y_0$ to achieve a good match of the two large sunspots and the several pores inside the SP FOV (lower left two panels of Figure \ref{fig1}). 

We then repeated the same procedure with the HMI vector field data at 12\,min cadence using the magnetic field strength of the HMI and SP to verify the match (e.g., upper left and lower right panel of Figure \ref{fig1}). The pseudo-scans used a few dozens intensity images at 45\,s cadence, but only the four 12\,min HMI vector data sets taken between UT 19:12 and UT 19:48. The alignment quality of the latter is still fully sufficient given that in the end all SP quantities were downsampled to the HMI spatial sampling of 0\farcs5 for the comparison.  
\section{Data Analysis  \label{sec_ana}}   
\subsection{Derivation of ${\bf B}$ for Hinode SP data}
To retrieve the magnetic field in the solar photosphere from the SP data, three different inversion approaches were considered. We first ran the Very Fast Inversion of the Stokes Vector \citep[VFISV;][]{borrero+etal2011} over the SP spectra. The VFSIV code assumes the simplifying Milne-Eddington (ME) approximation for the radiative transfer that is included as a source function that changes linearly with optical depth. A magnetic fill factor of 1 was used in this inversion and only the \ion{Fe}{i} line at 630.25\,nm was analyzed, which corresponds to the standard HMI approach. We then downloaded the ME results for SP data from the Community Spectro-polarimetric Analysis Center of the High Altitude Observatory\footnote{\url{https://www2.hao.ucar.edu/csac}}. Their code employs a stray light factor and a scaling between the two \ion{Fe}{i} lines at 630.15 and 630.25\,nm. Its results will be labeled "ME" inversion in the following. Finally, we used the Stokes Inversion based on Response functions code \citep[SIR;][]{cobo+toroiniesta1992} that includes the full radiative transfer equation assuming local thermodynamic equilibrium (LTE) ("SP SIR $B$" in the following). The SIR code can recover stratifications of physical parameters with optical depth. However, for simplicity and to be comparable to the ME inversions, we used only one node for the LOS velocity and all physical parameters of the magnetic field, i.e., the magnetic field vector is constant in optical depth and reflects the average value inside the formation height of the spectral lines \citep{weste+etal1998}. In this inversion, both spectral lines were analyzed and a stray light factor $\alpha$ was used but no complex model with multiple atmospheric components inside a single pixel with a relative fill factor \citep[e.g.,][]{beck+etal2008,beck+rezaei2009}. 

\subsection{Derived Quantities}
From a direct analysis of the data, we obtained the continuum intensity $I_c$ for HMI and SP and the maximal polarization degree $p(x,y)$ for the SP as
\begin{eqnarray}
p(x,y) = \mbox{max}\,\sqrt{Q^2+U^2+V^2}/I (x,y,\lambda)|_{\Delta\lambda}
\end{eqnarray}
in a small wavelength interval $\Delta\lambda$ around the core of the \ion{Fe}{i} line at 630.25\,mn, where $I$ is the intensity and $Q,U,$ and $V$ are the Stokes parameters that represent linear and circular polarization, respectively.

The inversion results then provide the magnetic field vector {\bf B}, the magnetic field strength $B$ in Gauss (G) and the stray light factor $\alpha$ where applicable. From the latter, we defined the magnetic fill factor that describes the area fraction inside a pixel that is filled with magnetic field as $f = 1 -\alpha$ since the stray light contribution mimics a field-free component inside the pixel. The LOS magnetic flux is given by the standard definition $\Phi_{LOS} = B \cos \gamma \,A$ with $\gamma$ the inclination of the magnetic field vector to the LOS and $A$ the area of a pixel, where $B$ is replaced by $B\,f$ for all results with a fill factor $f$. We defined the "effective total magnetic flux" as $\Phi_{eff} = B\,f$. This quantity will be labeled in G like $B$ in the following, but one should imagine it to be always implicitly multiplied by the area of a HMI or SP pixel. 

\subsection{Masking of Solar Surface Structures}
We determined the locations of different solar structures through a combination of thresholds in different quantities and manual identification. The locations of umbrae, penumbrae and pores inside the FOV were determined using thresholds in the continuum intensity with an additional constraint of having some minimal area (upper right panel in Figure \ref{fig1}). Plage and QS areas were marked manually by using rectangles inside the FOV. Those regions were then filtered by setting a threshold in the SP polarization degree to only retain pixels with $p>p_{lim}$ (top middle panel of Figure \ref{fig1}). The distinction between plage and QS was set by the presence or absence of extended connected areas with polarization signals. Because of the threshold in polarization degree, only locations with significant polarization signals in the SP data remain for the five types of separate surface structures (umbra, penumbra, pores, plage, and QS). The statistics for the whole FOV were derived from all pixels inside the FOV with $p>p_{lim}$.    

\subsection{Magnetic Field Extrapolation}
To test the effect of the upscaling of the HMI magnetic field strength on magnetic field extrapolations, we ran a potential field extrapolation and the non-force free (NFFF) extrapolation technique developed by \citet{hu&dasgupta2008soph} and \citet{hu+2008apj,hu+2010jastp} over the large SHARP cut-outs. 
In this method, the magnetic field $\mathbf{B}$ is constructed as follows:
\begin{equation}
\mathbf{B} = \mathbf{B_1}+\mathbf{B_2}+\mathbf{B_3}; \quad \nabla\times\mathbf{B_i}=\alpha_i\mathbf{B_i}
\label{e:b123}
\end{equation}
with $i=1,2,3$. The sub-fields $\mathbf{B}_i$ are linear force-free fields (LFFFs) with their respective constants $\alpha_i$. One can set $\alpha_1 \ne \alpha_3$ and $\alpha_2 = 0$ without loss of generality, which reduces $\mathbf{B_2}$ to a potential field. To find optimal values for the still undetermined pair $\alpha = \{\alpha_1, \alpha_3\}$, an iterative method is used that minimizes the average deviation between the observed ($\mathbf{B}_t$) and the calculated ($\mathbf{b}_t$) transverse field on the photospheric boundary. The deviation can be quantified by the metric $E_n$ \citep{prasad+2018apj} as 
\begin{equation}
E_n = \left(\sum_{i=1}^M |\mathbf{B}_{t,i}-\mathbf{b}_{t,i}|\times|\mathbf{B}_{t,i}|\right)/\left(\sum_{i=1}^M |\mathbf{B}_{t,i}|^2\right)
\label{en}
\end{equation}
where the sum runs over all $M$ grid points on the transverse plane. Weaker magnetic fields are suppressed by weighting the contribution of each grid point with its observed transverse field strength \citep[for more details see][]{hu&dasgupta2008soph,hu+2010jastp}.

The extrapolated field $\mathbf{B}$ is a solution to an auxiliary higher-curl equation:
\begin{equation}
\nabla\times\nabla\times\nabla\times\mathbf{B}+a_1\nabla\times\nabla\times\mathbf{B}+b_1\nabla\times\mathbf{B}=0.
\label{e:bnff2}
\end{equation}
This equation includes a second-order derivative $(\nabla\times\nabla\times\mathbf{B})_z=-\nabla^2 B_z$ at $z=0$, requiring vector magnetograms at two heights for calculating $\mathbf{B}$. To work with the available single-layer vector magnetograms, an algorithm developed by \citet{hu+2010jastp} was used. This algorithm introduces additional steps to iteratively correct the potential sub-field $\mathbf{B_2}$. Beginning with an initial guess $\mathbf{B_2} = 0$, the problem reduces to second order, allowing boundary conditions for $\mathbf{B_1}$ and $\mathbf{B_3}$ to be determined using the trial-and-error process described above. If the minimum $E_n$ value is unsatisfactory, a corrector potential field for $\mathbf{B_2}$ is derived from the difference in transverse fields, i.e., $\mathbf{B}_t - \mathbf{b}_t$, and added to the previous $\mathbf{B_2}$ to improve the match, as measured by $E_n$.

From the extrapolation results that provide the magnetic field vector {\bf B}$(x,y,z)$ in a 3D volume we derived continuous open and closed magnetic field lines that trace the connectivity using the VAPOR visualisation package \citep{li+etal2019}, the magnetic field strength $B$ and its gradient with height, the (free) magnetic energy, electric currents and the Lorentz force using their standard definitions \citep[see, e.g.,][]{gary+etal1995,gary2009}.

\section{Results \label{sec_res}} 
\begin{figure}
\resizebox{8.8cm}{!}{\includegraphics{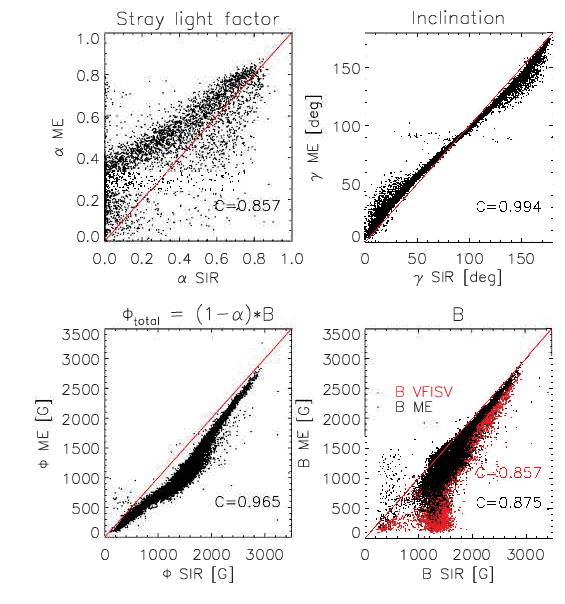}}
\caption{Scatterplots of quantities in the ME and SIR inversion of the SP data. Clockwise, starting left top: stray light factor $\alpha$, inclination $\gamma$, field strength $B$, and effective total flux $B\,f$. 
 The red lines indicate a unity slope. The linear correlation coefficients are given in the bottom right corner of each panel. The red dots in the lower right panel correspond to the VFISV ME inversion and were overplotted as second layer to enhance their visibility.}\label{fig2}
\end{figure} 
\begin{figure*}
\resizebox{17.6cm}{!}{\includegraphics{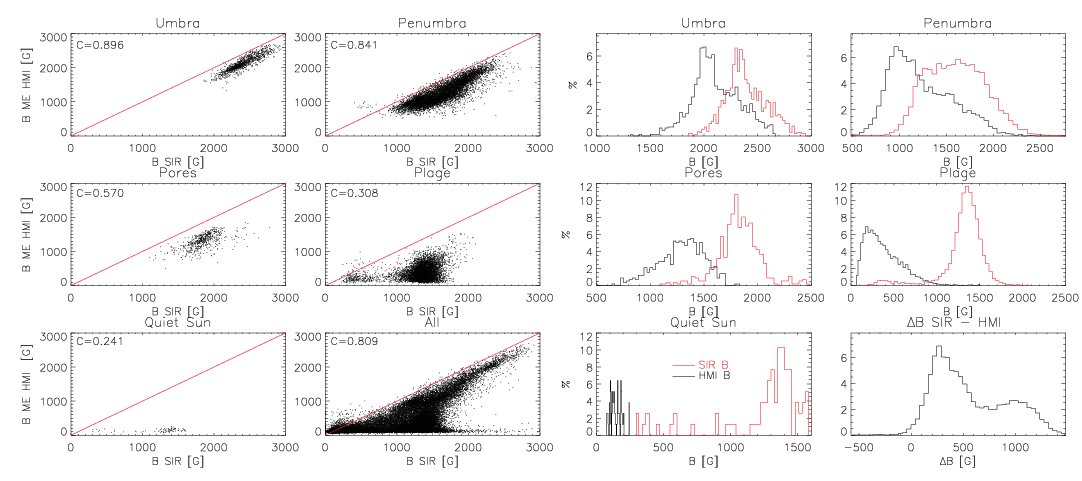}}
\caption{Scatterplots and histograms of field strength $B$ in different structures. First column, bottom to top: scatterplots between SP SIR $B$ and HMI $B$ for quiet Sun, pores, and umbra. Second column: for full FOV, plage and penumbra. The red lines indicate a unity slope. The linear correlation coefficients are given in the top left corner of each panel. Third column: histograms for quiet Sun, pores, and umbra for HMI $B$ (black lines) and SP SIR $B$ (red lines). Fourth column: histograms for the difference $\Delta B$ across the full FOV, and $B$ in plage and penumbra.}\label{fig3}
\end{figure*}     
\subsection{Inversion of SP Data}
We ran three different inversion approaches over the same SP data set, two of which included a fill factor $f$ (SIR, ME). Figure \ref{fig2} shows scatterplots of the magnetic field strength $B$, inclination $\gamma$, stray light factor $\alpha$ and effective magnetic flux $\Phi = B\,f$ over the full SP FOV between the inversion approaches. The SIR and ME results show a high correlation ($>0.85$) in all quantities. The largest differences are seen in $\alpha$, which could result from a trade-off between $\alpha$ and the different approaches to treat the radiative transfer in the inversion (ME vs.~LTE), which affects the intensity spectrum more than polarization. The results of the VFISV inversion with a fill factor of 1 are only overplotted for the field strength (lower right panel in Figure \ref{fig2}). They deviate prominently at low field strengths where the VFSIV $B$ stays at almost constant values of about 200--300\,G for the whole range of 200--1500\,G in the SP SIR $B$.

In the following, we will use the SIR inversion of the SP data as the best estimate for the true magnetic field properties since it represents the most realistic inversion setup (stray light factor $\equiv$ fill factor for unresolved magnetic fields, LTE radiative transfer, usage of both \ion{Fe}{i} lines). 

\begin{table*}
\caption{Average fill factor $f$ and field strength in different structures and all inversion approaches. All values apart from $f$ are in Gauss.}
\begin{tabular}{c|c|cccc|cc}
region &  $f$  &  SP SIR $B$& SP ME $B$ & SP VFISV $B$ & HMI $B$ & SP SIR $B\,f$ & HMI $B$ scaled \cr\hline\hline 
umbra    & 1.00 & 2405 & 2300 & 2122 & 2124 & 2403 & 2431 \cr
penumbra & 0.99 & 1623 & 1351 & 1246 & 1253 & 1609 & 1689 \cr
pores    & 0.99 & 1847 & 1679 & 1373 & 1303 & 1839& 1741 \cr
plage    & 0.61 & 1329 & 1143 & 510  & 394 & 818 & 646 \cr
QS       & 0.46 & 1247 & 1034 & 282 &  145 & 540 & 147  \cr
\end{tabular}\label{tab1}
\end{table*}

\begin{figure*}
\resizebox{17.6cm}{!}{\includegraphics{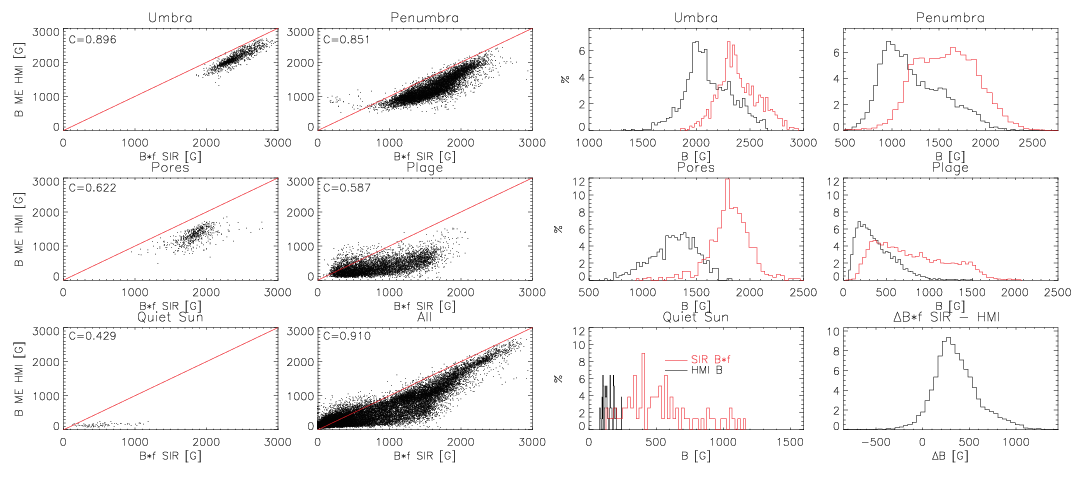}}
\caption{Scatterplots and histograms of field strength $B$ in HMI and total effective flux $B\,f$ in SP data in different structures. Same layout as Figure \ref{fig3}.}\label{fig5}
\end{figure*} 
\subsection{Field Strength SP SIR $B$ vs.~HMI $B$}
Figure \ref{fig3} compares the field strength $B$ in the SP SIR results and the HMI data for the different type of structures inside the FOV defined above through scatterplots (left two columns) and histograms (right two columns). The values match for locations in the umbra and penumbra with a high correlation ($>0.8$). The corresponding histograms of SP and HMI (top right panels of Figure \ref{fig3}) have a similar shape with some global offset with a higher $B$ in the SP data. For pores (middle row), the correlation drops to about 0.57 with an increasingly larger offset in the histograms. For plage locations, the scatterplot shows no clear visual correlation anymore, with the HMI results clustered at a nearly constant value of $B <500$\,G below the SP values, similar to the behavior of the VFSIV SP results without $f$ in Figure \ref{fig2}. The shape of the histograms of $B$ for the two approaches does not match for plage regions with a dominant Gaussian at about 1.3\,kG for the SP SIR while HMI $B$ peaks near 0.2\,kG with a weak tail towards higher field strengths. The mismatch gets even worse for QS regions, where the HMI data exhibits a constant value of $B<200$\,G (lower left panel of Figure \ref{fig3}), while SP SIR $B$ reaches up to 1.5\,kG. The scatterplot for the full FOV (bottom panel in second column) summarizes the findings: a reasonable match of SP SIR and HMI for $B>1.5$\,kG and a clear mismatch otherwise, where all HMI values are significantly lower than the SP results at a nearly constant value. The same is seen in the histogram of the field strength difference (lower right panel of Figure \ref{fig3}) across the full FOV that shows a bi-modal distribution with one roughly Gaussian peak at $+300$\,G from umbra, penumbra and pores, and a second peak at $+1$\,kG that reflects the plage and QS areas.  

The average values of $B$ and $f$ in the different structures and all inversion approaches are listed in the first five columns of Table \ref{tab1}. They confirm the clear distinction: wherever the fill factor $f$ drops from unity (plage, QS), the inversion approaches without $f$ (VFSIV, HMI) give significantly lower values of $B$. The factor for plage is 2--3 and increases to 5--10 in the QS. For umbra, penumbra and pores, the values for HMI and VFISV are 200--500\,G lower than for the SP SIR and ME, but stay above 1.2\,kG. 

\subsection{Effective Total Flux SP SIR $B\,f$ vs.~HMI $B$}
The absence of a fill factor in the inversion of spatially unresolved magnetic fields showed a strong effect on the field strength $B$ in the previous section. We thus decided to compare the effective total magnetic flux $B\,f$ in the SP SIR inversion with the field strength $B$ in HMI as the next step. Figure \ref{fig5} shows scatterplots and histograms of those two quantities in the same layout as Figure \ref{fig3}. As expected, locations with a large fill factor close to unity (umbra, penumbra and pores) show only minor changes in both scatterplots and histograms. The correlation for pores increases slightly to about 0.6. For both plage and QS areas, the correlation, however, nearly doubles to 0.587 and 0.429, respectively. The shape of the plage histogram for the SP SIR $B\,f$ significantly differs from SP SIR $B$ alone and now resembles the one of HMI $B$ with a maximum at low field strengths much better apart from a scaling factor in the modulus. The histogram for SP SIR $B\,f$ in QS is now more compact at lower $B\,f$ values of about 500\,G. In the histograms of the difference between SP SIR $B\,f$ and HMI $B$ (lower right panel) the second peak at about 1\,kG has disappeared leaving a single Gaussian at $+300$\,G. The comparison of the scatterplots for the plage regions and the full FOV (second column) in Figures \ref{fig3} and \ref{fig5} clearly demonstrates that this is caused by the better match of plage areas. To first order, SP SIR $B\,f$ and HMI $B$ show visually a clear correlation over the full range of field strength or total flux values with a correlation coefficient of 0.91 (bottom panel in the second column of Figure \ref{fig5}).  
\begin{figure}
\centerline{\resizebox{7.cm}{!}{\includegraphics{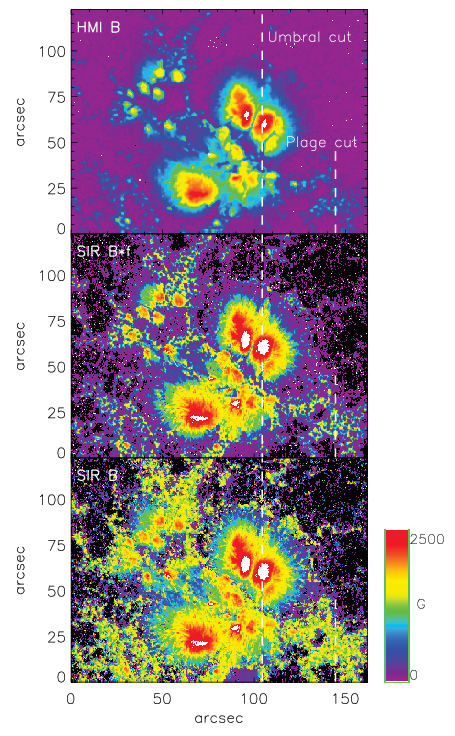}}}
\caption{2D maps of field strength SIR $B$ (bottom panel) and total effective flux SIR $B\,f$ in SP data (middle panel), and field strength $B$ in HMI data (top panel). The two vertical white dashed lines indicate the locations of spatial cuts across an umbra and a plage region shown in Figure \ref{fig8}.}\label{fig7}
\end{figure}   

To better understand this behavior, we looked in more detail at the differences in the results for plage areas. Figure \ref{fig7} shows 2D maps of $B$ for SP and HMI and $B\,f$ for the SP. The map of SP SIR $B$ shows clear "blooming" around network and plage elements, i.e., the value of $B$ stays nearly constant at $> 1$\,kG over a distance of a few arcsec from the location of each magnetic element (bottom panel of Figure \ref{fig7}). The locations in the QS (upper right corner of the FOV at $x\sim 150^{\prime\prime}, y\sim 85^{\prime\prime}$) have kG fields in the SIR inversion, while they are nearly invisible in the HMI $B$ map (top panel of Figure \ref{fig7}). The spatial pattern around plage and network elements changes drastically for SP SIR $B\,f$ (middle panel of Figure \ref{fig7}). The "blooming" disappears and the value of the total effective flux reduces smoothly with the distance from plage and network elements, which makes the spatial patterns in SP SIR $B\,f$ much more similar to those in HMB $B$. All areas with a high fill factor (umbra, penumbra, pores) are rather similar in all three panels and quantities.
\begin{figure}
\resizebox{8.8cm}{!}{\includegraphics{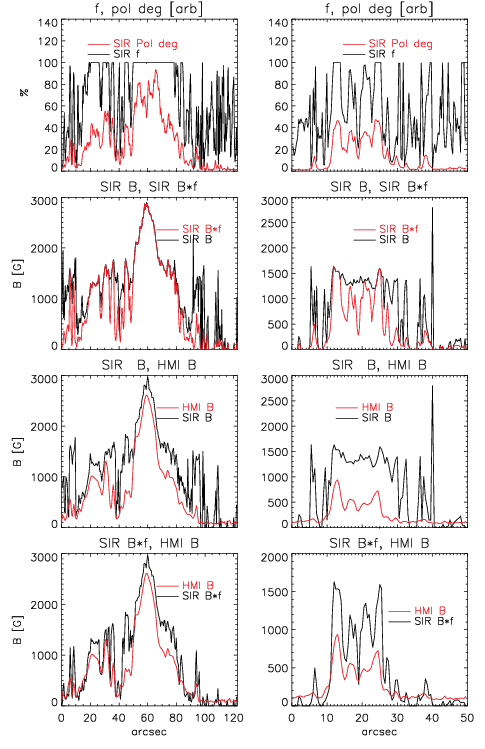}}
\caption{Different quantities along the spatial cuts across an umbra (left column) and a plage region (right column) marked in Fig.~\ref{fig7}. Top to bottom: fill factor $f$ (black lines) and polarization degree $p$ (red lines), field strength $B$ (black lines) and effective total flux $B\,f$ (red lines) in SP data, field strength in SP (black lines) and HMI data (red lines), and $B\,f$ in SP data (black lines) and $B$ in HMI data (red lines).}\label{fig8}
\end{figure}   

The fact that the fill factor is the controlling parameter is visualized by the cuts through one of the umbrae and a plage region that are shown in Figure \ref{fig8}. For all locations with $f\sim 1$ (left column, $x\sim 50^{\prime\prime}-80^{\prime\prime}$), SP SIR $B$, SP SIR $B\,f$ and HMI $B$ are very similar. The cut through the plage region in the right column shows that the true field strength $B$ in the SIR inversion is a rather constant 1.5\,kG in the plage region (middle two panels in the right column, $x\sim 10^{\prime\prime}-27^{\prime\prime}$). In the value of SP SIR $B\,f$, one can detect instead four clearly separate structures over the same spatial range, each one representing a magnetic element and its immediate surroundings which can also be identified in the 2D maps of Figure \ref{fig7} along the cut through the plage. In the SP SIR $B\,f$, these separate features are just the result of the variation of the value of $f$ (top right panel of Figure \ref{fig7}), which itself can be traced back to the variation of the polarization degree $p$. The bottom panel of the right column of Figure \ref{fig8} shows that the "field strength" of HMI in plage is in reality sampling the total effective flux $B\,f$ in the SIR inversion rather than the constant field strength SP SIR $B$. 

The different behavior will be discussed in more detail below, but to first order it is not too surprising. In the SP data, pixels around magnetic elements have about the same polarization signal in the spectral dimension as at its center with just a reduced amplitude because of the spatial PSF that dilutes the signal into the immediate surroundings. The SIR inversion can successfully determine the field strength $B$ in all cases and uses the fill factor to match the polarization amplitude, which causes the "blooming" effect in SP SIR $B$. For HMI (or VFSIV $B$), the analysis approach is instead forced to reduce the field strength with increasing distance to network elements to reproduce the observed decreasing polarization amplitudes. 

Since there is no way to match HMI and SP SIR $B$ without using a fill factor or similar parameter and the magnetic field extrapolation code would be unable to deal with that, we thus decided to match HMI $B$ and SP SIR $B\,f$ as the best possible compromise.   
\begin{table*}
\caption{Average ratio and correlation coefficient $C$ between different quantities.}
\begin{tabular}{c|cc|cc|cc}
region & B(SP SIR)/B(HMI) & $C$ & $B\,f$ (SP SIR)/B(HMI) & $C$ & $B\,f$ (SP SIR)/B(HMI) scaled & $C$ \cr\hline\hline
umbra & 1.14 & 0.896 & 1.14 & 0.896 & 0.99 & 0.882\cr
penumbra & 1.33 & 0.841 & 1.31 & 0.851 & 0.95 & 0.851\cr
pores & 1.45 & 0.570 & 1.44 & 0.622 & 1.06 & 0.621\cr
plage & 4.48 & 0.308 & 2.43 & 0.587 & 1.82 & 0.587\cr
quiet Sun & 9.19 & 0.241 & 3.83 & 0.429 & 3.81 & 0.413\cr
full FOV & 2.53 & 0.809 & 1.73 & 0.910 & 0.94 & 0.903\cr 
\end{tabular}
\label{tab2}
\end{table*}
\begin{figure}
\centerline{\resizebox{8.cm}{!}{\includegraphics{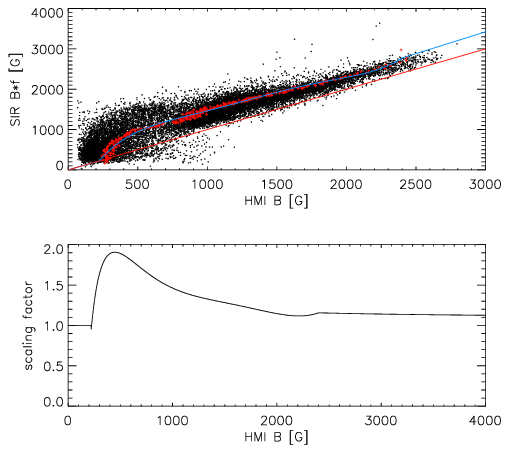}}}
\caption{Scatterplot of HMI $B$ and SP SIR $B\,f$ (top panel) and scaling curve for HMI data derived from it (bottom panel). The red pluses in the top panel are binned data points, the blue line a polynomial fit and the red solid line is at unity slope.}\label{fig10}
\end{figure}
  
\subsection{Determination of Scaling of HMI $B$ to SP SIR $B\,f$}
We used a scatterplot between HMI $B$ and SP SIR $B\,f$ to determine a scaling curve for the HMI field strength (Figure \ref{fig10}). Only pixels with a significant polarization signal in the SP data were considered ($p>p_{lim}$, see Figure \ref{fig1}). We determined the average values of SP SIR $B\,f$ for bins in HMI $B$ (red pluses in the top panel of Figure \ref{fig10}) and fitted a 5th-order polynomial to the binned values. As there are little to no pixels with HMI $B < 220$\,G above the threshold $p_{lim}$ in SP or with $B > 2400$\,G, we replaced the polynomial curve with unity for HMI $B <220$\,G, i.e., the original values of HMI $B$ in that range are not modified, and extended the polynomial by a straight line for HMI $B>2400$\,G (bottom panel of Figure \ref{fig10}).  The slope to be used at the upper end of the scaling curve was taken from the corresponding value of the polynomial around 2400 G. It might be slightly too large, but only very few umbral pixels are affected that generally lead to open field lines leaving through the upper boundary of the extrapolation box. At the lower end, the rms noise of HMI $B$ on locations without significant polarization signal in the SP data is about 100\,G, which we initially used as the limit for changing the scaling to unity to avoid noise amplification. In the end, we switched to a more conservative threshold of 220\,G because otherwise the application of the scaling curve led to a clear "salt'n'pepper" noise pattern in the upscaled HMI data, especially for the large FOV of the extrapolation. As discussed below in Section \ref{lim_scaling}, one would need a measure of the polarization degree in the HMI data instead to better distinghuish between genuine and spurious values as the modulus of the HMI field strength alone does not work reliably. The scaling value of unity was selected to ensure that any signals that might be genuine in the HMI data stay at their original values. The scaling curve thus leaves HMI $B$ values below 220\,G untouched, rises to an upscaling of about 2 for HMI $B\sim 400-500$\,G and decreases slowly to slightly above 1 again at 2\,kG.

\begin{figure}
\resizebox{8.8cm}{!}{\includegraphics{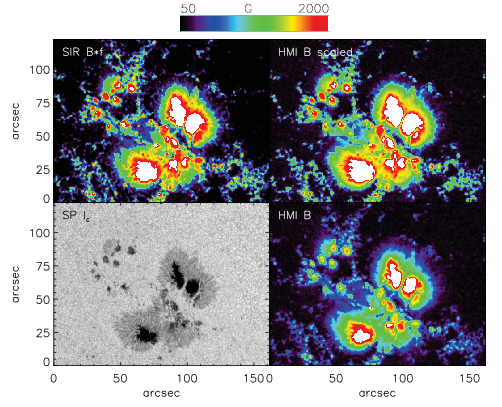}}
\caption{2D maps of clockwise, starting left bottom: continuum intensity $I_c$ and total effective flux $B\,f$ from SP, and upscaled and original field strength $B$ from HMI.}\label{fig11}
\end{figure}

\begin{figure*}
\resizebox{17.6cm}{!}{\includegraphics{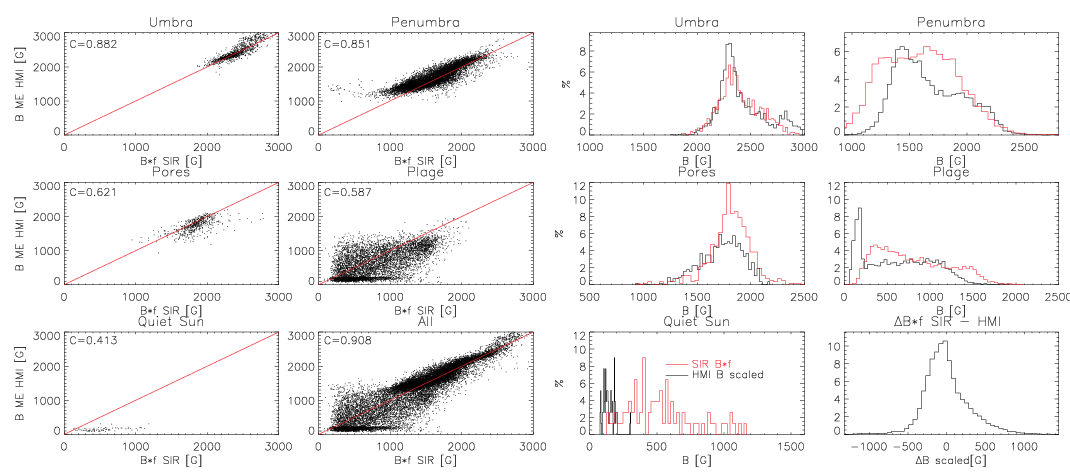}}
\caption{Scatterplots and histograms of $B\,f$ in SP and upscaled HMI $B$ data. Same layout as in Fig.~\ref{fig3}.}\label{fig12}
\end{figure*}
\begin{figure}
\centerline{\resizebox{8.cm}{!}{\includegraphics{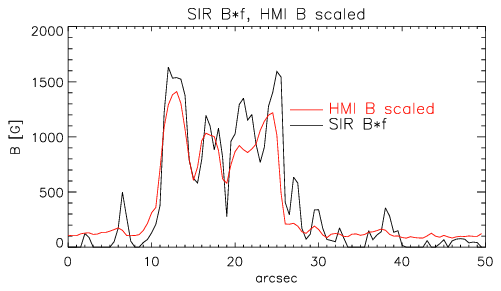}}}
\caption{Effective total flux $B\,f$ in SP and upscaled HMI $B$ along the spatial cut through a plage region marked in Figure \ref{fig7}.}\label{fig13}
\end{figure}
\subsection{Upscaling of HMI $B$}
For the application, we read off the field strength value HMI $B(x,y)$ for each pixel and multiplied it with the scaling factor associated to that field strength in the curve. After the application, the average ratio between the scaled and original HMI $B$ was about 1.1 in the umbrae and 1.6 in the plage regions. Figure \ref{fig11} shows the 2D maps of HMI $B$, SP SIR $B\,f$, the scaled HMI $B$ and a continuum intensity image as reference. The main difference between the original and scaled HMI $B$ maps is the clear enhancement of the field strength in the plage and QS regions, while the umbra, penumbra and pore regions changed only slightly. The upscaled HMI $B$ matches the spatial patterns and modulus of the SP SIR $B\,f$ map well (left and right panel in top row of Figure \ref{fig11}).

To quantify the improved match, Figure \ref{fig12} shows the same  scatterplots and histograms as used before for SP SIR $B\,f$ and the upscaled HMI $B$. All correlation values stayed about the same as for the original HMI $B$ and SP SIR $B\,f$. For the umbra, penumbra and pore regions the offset in field strength has been successfully removed in comparison to Figure \ref{fig5} (upper right panels of Figure \ref{fig12}), the histograms now overlap well. The strongest fields in the umbra with $B > 2700$\,G are slightly overamplified after the scaling, which again indicates that the value at the upper end of the scaling curve is somewhat too high. The histograms for the penumbra show a slightly different shape, but cover the same range in $B$. The data points for the plage regions (middle row of 2nd column of Figure \ref{fig12}) now scatter around the line of unity correlation, but with a comparable large range of up to 1\,kG difference between SP SIR $B\,f$ and the upscaled HMI $B$ value. The histograms for plage regions now cover the same range in $B$ opposite to before. The peak in the upscaled HMI $B$ histogram for plage at $<220$\,G results from all pixels that were not modified, about 8\,\% of the plage area. The scatterplot for the full FOV now shows a correlation around unity for all field strengths, while the difference of the upscaled HMI $B$ and SP SIR $B\,f$ yields a single Gaussian distribution that is roughly centered at zero (bottom right panel of Figure \ref{fig12}).

\begin{figure*}
\centerline{\resizebox{16.cm}{!}{\includegraphics{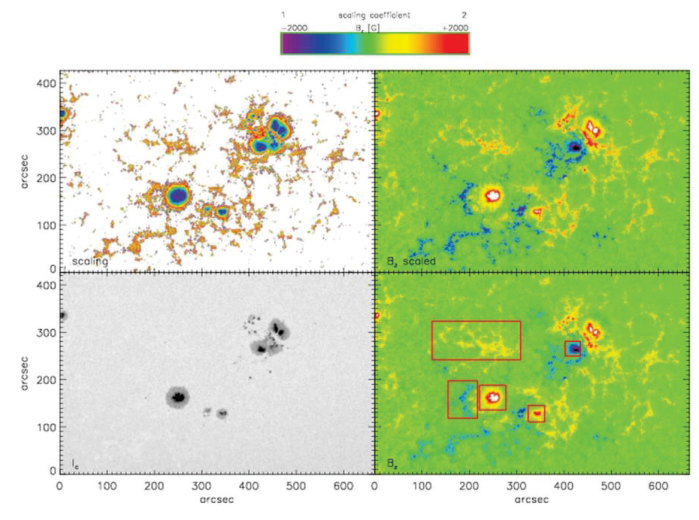}}}
\caption{2D maps of the FOV of the extrapolation box. Left column: continnum intensity $I_c$ (bottom panel) and scaling factor (top panel). Right column: original (bottom panel) and upscaled (top panel) vertical magnetic field $B_z$ from HMI. The red rectangles in the lower right panel indicate the locations of seed points of magnetic field lines for checking up on changes of connectivity in plage and sunspots.}\label{fig14}
\end{figure*}

Figure \ref{fig13} visualizes the improvement of the match along the same cut through the plage region as in Figure \ref{fig7}. The upscaled HMI $B$ and SP SIR $B\,f$ now have the same spatial patterns and about the same amplitudes.

The last two columns of Table \ref{tab1} list the average values of SP SIR $B\,f$ and HMI $B$ after the upscaling, while Table \ref{tab2} has the average ratios and correlation coefficients between different quantities. For the original HMI $B$ and SP SIR $B$, HMI $B$ underestimates the field strength on average by a factor of 4.48 in plage and 9.19 in QS, while for structures with $f\sim 1$ (umbra, penumbra, pores) the factor is $<1.5$. The correlation values for plage and QS are $<0.5$. Using SP SIR $B\,f$ instead increases the correlation for plage, QS and the full FOV and reduces the ratios to 2.43 for plage and 3.83 for QS. The upscaling of HMI $B$ leaves the correlations untouched, but now brings the ratios for umbra, penumbra and pores close to unity and for plage to about 1.82. Without some additional filtering for locations with significant polarization signal in HMI instead of SP and an improved scaling for genuine HMI $B$ values $<220$\,G it seems impossible to achieve any better match. 

\begin{figure}
\resizebox{7.5cm}{!}{\includegraphics{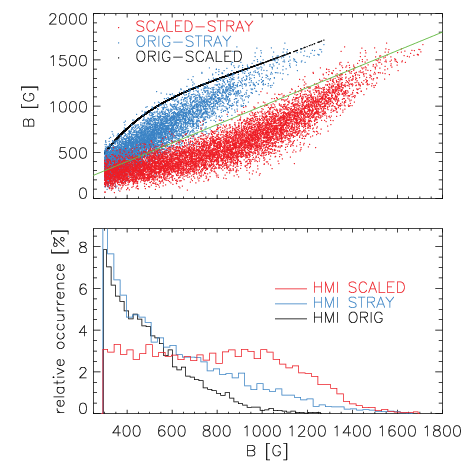}}
\caption{Comparison of the field strength $B$ in original, scaled and stray-light corrected HMI data for plage regions. Top panel: scatter plots of $B$ in original vs.~scaled data (black dots), original vs.~SLC data (blue dots) and scaled vs.~SLC data (red dots) in plage. The green line indicates unity correlation. Bottom panel: histograms of $B$ for original (black line), SLC (blue line) and scaled HMI data (red line).} \label{fig_stray_corr}
\end{figure}

\subsection{Magnetic Field Extrapolation}
\subsubsection{Application of HMI Upscaling to Extrapolation Box}
The FOV of the SP scan covers both polarities of NOAA 12104, but is still comparably small for an extrapolation. The scaling curve does not require to stay restricted to it since its applictation is only based on the value of HMI $B$. We thus applied the scaling curve to a large cut-out from HMI that covers both NOAA 12104 and NOAA 12107 instead. The scaling is only applied to the magnetic field strength and leaves the field orientation untouched. Figure \ref{fig14} shows the FOV used for the extrapolation prior and after the upscaling of HMI. The primary effects are the same as before for the SP FOV, a slight enhancement in sunspots and a strong enhancement in plage areas. The map of the scaling factor (upper left panel in Figure \ref{fig14}) naturally shows the same spatial pattern as the initial field strength, but with a clear difference in modulus inside ($\sim 1$) and outside ($\sim 2$) of sunspots. 

\subsubsection{Comparison to Stray-Light Corrected HMI Data}
For the FOV used in the extrapolation, we also have stray-light corrected (SLC) HMI data available (courtesy A. Norton). The major effect of our upscaling is in QS surroundings, so we only used the plage region sample for a comparison. Figure \ref{fig_stray_corr} shows scatterplots and histograms of the magnetic field strength $B$ in the original, scaled and SLC HMI data for plage regions in the large FOV. Our scaling curve to match HMI $B$ with the effective magnetic flux $B\,f$ in the SP data turns out to be the upper envelope of the SLC data points (black and blue dots in the upper panel of Figure \ref{fig_stray_corr}). The stray-light correction increases the field strength, but to a lesser amount than our correction curve, e.g., almost all red dots in the scatter plot of scaled vs.~SLC data are below the unity line. The same weaker enhancement in the SLC data is also clearly seen in the histograms in the bottom panel of Figure \ref{fig_stray_corr}: in the SLC data, the field strength in plage still stays far below 1\,kG for most locations with the maximum of the distribution at low field strengths. The SLC approach thus also falls short of the effective flux $B\,f$ in the SP, which itself is significantly lower than the actual "true" field strength SIR $B$.        

We thus decided to only run both a potential and the NFFF magnetic field extrapolation over the original and upscaled HMI data without considering the SLC data further on.
\subsection{Effects on Magnetic Field Extrapolation}
\begin{figure*}
\centerline{\resizebox{17.cm}{!}{\includegraphics{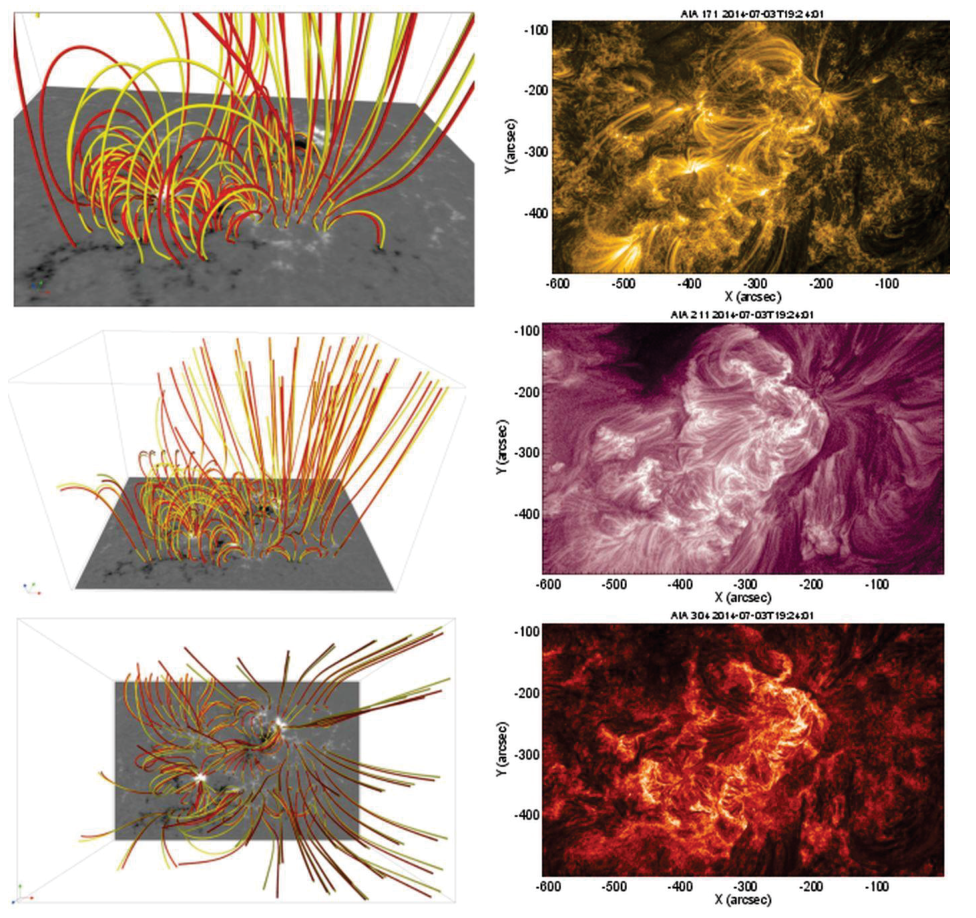}}}
\caption{3D views of the magnetic field extrapolation (left column) and concurrent AIA images (right column). Left column, bottom to top: top view of the extrapolation box with magnetic field lines from the same seed points before (red) and after scaling of the HMI magnetogram (yellow), side view (middle panel) and zoom on field lines of one spot and plage region (top panel). The background image shows HMI $B_z$. Right column, bottom to top: AIA images at 304, 211, and 171 {\AA}.}\label{fig_3d} 
\end{figure*}

\begin{table}
\caption{Number of open and closed field lines.}
\begin{tabular}{c|cccc}
type & \# seed points & open & closed & \% closed \cr\hline
\multicolumn{5}{c}{original HMI magnetogram}\cr\hline
full FOV & 10000& 3359& 6641 & 66\cr
sunspots & 7500 & 556 & 6944 & 93\cr
plage & 5000 & 1060& 3940 &  79 \cr\hline\hline
\multicolumn{5}{c}{scaled HMI magnetogram}\cr\hline
full FOV & 10000&3406 & 6594& 66\cr
sunspots & 7500 & 360 & 7140  & 95 \cr
plage & 5000 & 1143& 3857 &  77 \cr
\end{tabular}
\label{tab_fieldlines}
\end{table}

\subsubsection{Magnetic Connectivity}
Figure \ref{fig_3d} shows 3D visualizations of the magnetic field extrapolation together with concurrent AIA images. Both the extrapolation and the AIA images exhibit primarily open field lines to the South and West of the ARs. Between the two ARs and from their sunspots to the plage regions a mixture of open and closed field lines is seen. Most of the field lines from the original HMI extrapolation (red lines) and after the upscaling (yellow lines) match very closely.  
At many places where the connectivity visually changed, the outer end point of closed loops often only just moved to a close-by point in the same plage region. Closed field lines with a significant change of connectivity after the upscaling have a tendency to form taller loops (top left panel of Figure \ref{fig_3d}), with some of them now leaving the extrapolation box through the sides to close outside of the FOV. 


Table \ref{tab_fieldlines} lists the number of open and closed field lines for the full FOV and the sunspot and plage regions indicated in Figure \ref{fig14}. The majority of field lines ($\sim 66$\,\%) is closed in both extrapolations. The difference in the number of closed field lines between the two extrapolations is small at a level of 2\,\% for all three samples with only a weak trend for more closed field lines for the sunspots. The global pattern of connectivity was thus not significantly changed by the upscaling of the HMI field strength.

\subsubsection{Magnetic Field Line Properties: Height, Length, Horizontal Distance}
\begin{figure}
\resizebox{8.cm}{!}{\includegraphics{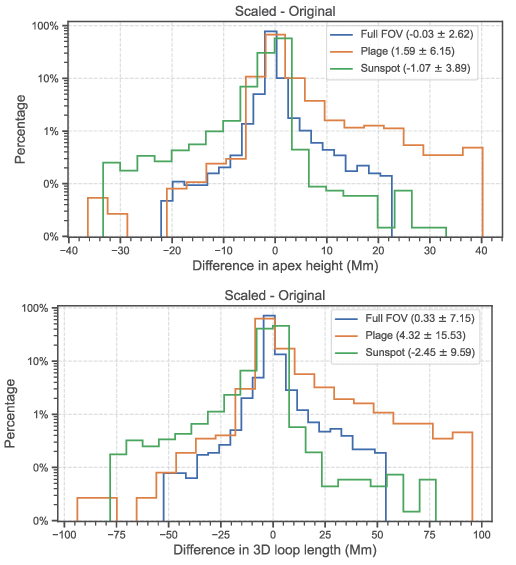}}
\caption{Difference of the apex height (top panel) and the 3D loop length (bottom panel) of closed loops before and after the scaling. Blue lines: full FOV. Orange lines: plage. Green lines: sunspots. Median values in Mm are given at the upper right corner. The y-axes are on a logarithmic scale.}\label{fig_height} 
\end{figure}
\paragraph{Apex Height of Closed Loops} To quantify the change of closed field lines by the upscaling, we determined the apex height for all seed points that generated closed loops in both extrapolations. The top panel of Figure \ref{fig_height} shows the difference of the apex height (after scaling minus original) for the full FOV, plage and sunspot sample. The apex height of the majority of the closed field lines changed by less than 10\,Mm. The plage regions show some increase in apex height for about 10\,\% of the sample with an average increase of 1.6\,Mm, while the sunspot sample shows the opposite trend with a few percent somewhat lower apex heights.   
\paragraph{3D Length of Closed Loops} The bottom panel of Figure \ref{fig_height} shows the difference in the 3D length of closed field lines, i.e., the total path length along closed field lines, in the same layout. The picture is very similar with changes of less than 20\,Mm in length in most cases, and an about 10\,\% fraction of plage (sunspot) loops that are significantly longer (shorter) than before.

\begin{figure}
\resizebox{8.cm}{!}{\includegraphics{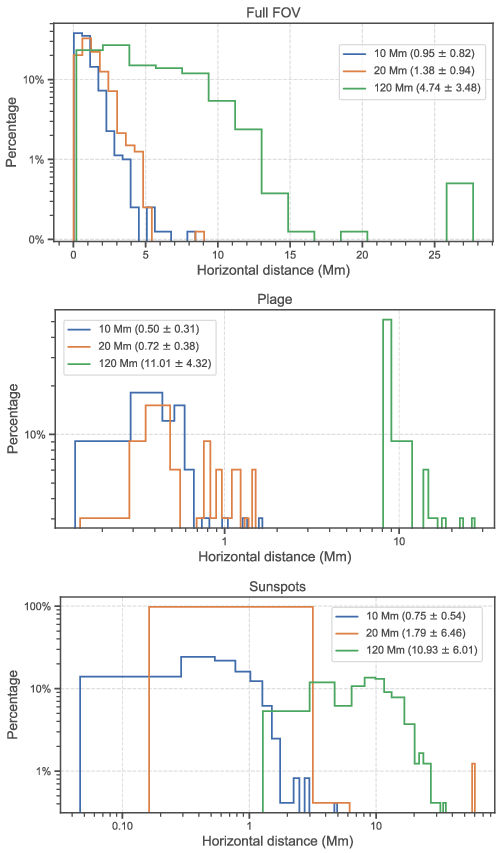}}
\caption{Horizontal distance between open field lines from the same seed point before and after the scaling at 3 heights of 10 (blue lines), 20 (orange lines), and 120\,Mm (green lines). Top panel: full FOV. Middle panel: plage. Bottom panel: sunspots.}\label{fig_open_lateral} 
\end{figure}

\paragraph{Horizontal Distance Between Field Lines} Figure \ref{fig_open_lateral} shows the lateral horizontal distance between open magnetic field lines from the same seed point in both extrapolations. We determined the distance at three different heights of 10, 20, and 120\,Mm. For heights up to 20\,Mm, the distance stays well below 10\,Mm for most cases, while at a height of 120\,Mm the average horizontal distance is about 10\,Mm. Again only a small percent fraction changed by larger distances of 20\,Mm or more.    
 
\begin{figure}
\resizebox{8.5cm}{!}{\includegraphics{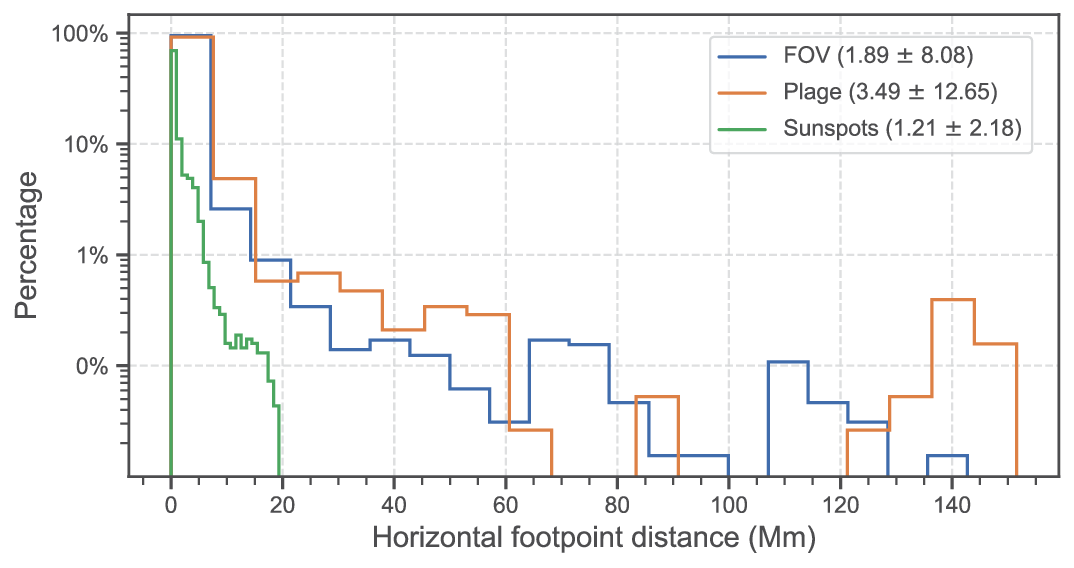}}
\caption{Horizontal distance between the outer footpoints of closed field lines from the same seed point before and after the scaling. Blue line: full FOV. Orange line: plage. Green lines: sunspots. Median values in Mm are given at the upper right corner.}\label{fig_closed_lateral} 
\end{figure}

For closed magnetic field lines, we calculated the horizontal distance of the outer footpoints of closed loops that started from the same seed point (Figure \ref{fig_closed_lateral}). That graph maybe shows the pattern in the most direct way. Only a small percent fraction of the outer footpoints of closed field lines in plage or the full FOV changed by more than 20\,Mm, while none of the closed field lines starting from inside a sunspot exceeded that value.   

In total, changes in apex height, length, and horizontal distance for open or closed field lines stay below about 10\,Mm from the botttom layer to a height of about 20\,Mm for the large majority of the field lines, which implies again only a small fraction where the connectivity significantly changed by the upscaling.  

\begin{figure}
\resizebox{8.8cm}{!}{\includegraphics{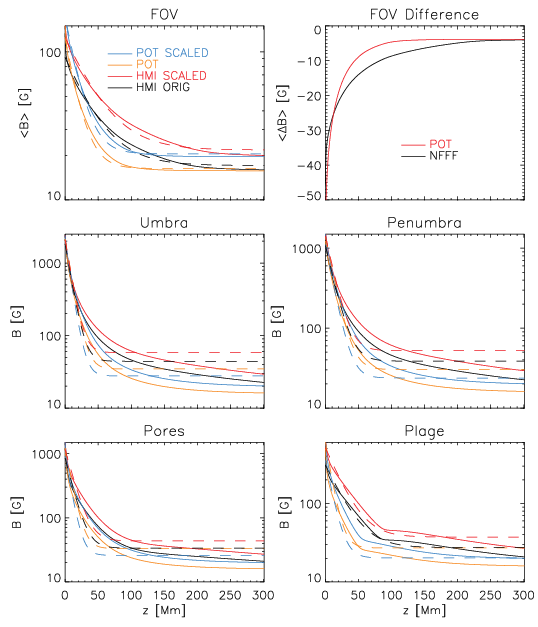}}
\caption{Average field strength with height $B(z)$ for different structures. Left column, bottom to top: average $B(z)$ for pores, umbra and the full FOV. Right column, bottom to top: average $B(z)$ for plage and penumbra, and difference of average $B(z)$ between the original and scaled HMI data across the full FOV. Black (red) lines: original (scaled) HMI data in the NFFF extrapolation. Orange (blue) lines: original (scaled) HMI data in the potential extrapolation. Dashed lines in the same colors: exponential fit.}\label{fig15}
\end{figure}
\subsubsection{Magnetic Field Strength}
For the extrapolation box with its larger FOV, we defined only four different samples. A mask of umbra, penumbra and pores was defined as before using intensity thresholds. For the plage, we selected the corresponding regions inside the FOV and considered only locations with $B>300$\,G in the original HMI data prior to the upscaling. This yielded a similar mask as in Figure \ref{fig1}, but precluded to define a QS sample as the values of HMI $B$ in the QS are often below the threshold. The QS sample is thus to some extent only represented by the average over the full FOV that was done without any additional constraint. 

Figure \ref{fig15} shows the average magnetic field strength with height $B(z)$ for the different cases. We fitted an exponential decay
\begin{eqnarray}
B(z) = B_0 \exp^{-\frac{z}{\sigma}} + C  \label{exp_drop}
\end{eqnarray}
to each curve, where $z$ is the height in Mm, $\sigma$ the scale height and $C$ the value at the upper boundary in height. Table \ref{tab1a} lists values of $B$ in steps of 30\,Mm, while Table \ref{tab1b} has the scale heights of the magnetic field strength in its last two columns. 

\begin{table*}
\caption{Magnetic field strength with height $B(z)$ in G in the NFFF extrapolation. The bottom two rows give values from radio measurements for comparison.}
\begin{tabular}{c|cccccccccc}
$z$ [Mm] & 0 & 30 & 60 & 90 & 120 & 150 & 180 & 210 & 240 & 270\cr\hline\hline
full FOV & 134 & 48 & 32 & 25 & 21 & 19 & 17 & 17 & 16 & 16\cr
full FOV scaled & 179 & 67 & 45 & 34 & 29 & 25 & 23 & 21 & 21 & 20\cr
umbra & 2117 & 169 & 75 & 50 & 40 & 35 & 31 & 28 & 26 & 24\cr
umbra scaled & 2429 & 211 & 99 & 67 & 53 & 46 & 41 & 37 & 34 & 21\cr
plage & 506 & 112 & 55 & 34 & 33 & 31 & 28 & 26 & 24 & 22\cr
plage scaled & 881 & 166 & 79 & 46 & 44 & 41 & 37 & 34 & 31 & 29\cr\hline
approximate $z$ [Mm] & 15 & 38 & 50 & 90 & 110 & 130&-- & 200 & --& 300 \cr\hline\hline
\citet{alissandrakis+gary2021}, Table 1 & 110 & 26--125 & 30--65 & 10--20 & 16--20 & 10--16 &-- & 10--15 &-- & 5\cr 
\end{tabular}\label{tab1a}
\end{table*}

The curves of $B(z)$ in Figure \ref{fig15} and the corresponding values in Table \ref{tab1a} show the behavior expected from the actual upscaling. There are only small changes for umbra, penumbra and pores and a larger difference over plage areas. At $z=0$\,km, the average ratio between the scaled and original HMI data was 1.74 for plage, 1.15 for umbrae, and 1.3 for the full FOV as determined after the application of the upscaling to the larger FOV of the extrapolation box. The $B(z)$ values above the regions then just follow this difference at the bottom boundary. 

At a height of about 150\,Mm the differences between original and scaled HMI data are $<10$\,G regardless of the structure at the photospheric boundary, while on average over the full FOV the difference is 40\,G at $z=0$\,km. The field strength value levels off at about 16--20\,G at the upper boundary of the extrapolation box at $z=300$\,Mm.

The scale height of the magnetic field strength is 20--40\,Mm for the full FOV and plage regions, but reduces to about 10\,Mm above all strong magnetic field concentrations (Table \ref{tab1b}). The potential field extrapolation drops twice as fast over the full FOV and plage region, but has the same magnetic scale height above sunspots and pores. The fits of the exponential in Figure \ref{fig15} show that the assumption of a constant magnetic scale height is not valid above strong magnetic field concentrations, where the decrease in $B$ slows down with height leading to an increase in the scale height. For the average over the full FOV (upper left panel in Figure \ref{fig15}) a single scale height provides a good match to the values.      

\begin{figure}
\resizebox{8.8cm}{!}{\includegraphics{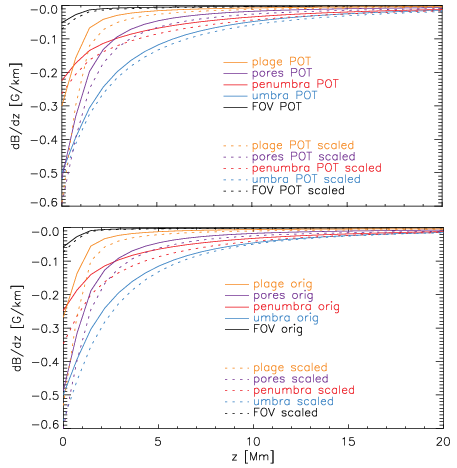}}
\caption{Gradient of magnetic field strength $dB/dz$ for different structures. Bottom panel: NFFF extrapolation. Top panel: potential field extrapolation. Black: full FOV. Blue: umbra. Red: penumbra. Purple: pores. Orange: plage. The solid (dashed) lines show the original (scaled) HMI data.}\label{fig17}
\end{figure}

Figure \ref{fig17} and the left two columns of Table \ref{tab1b} show the magnetic field strength gradient $dB/dz(z)$. The only difference between scaled and original HMI data is seen for plage areas, with an increase by about a factor of 2 from $-0.27$ to $-0.53$\,G\,km$^{-1}$ at $z=0$\,km. The gradients at $z=0$\,km range from -0.06 to -0.50\,G\,km$^{-1}$ and decrease with height. The potential and NFFF extrapolation have very similar values. The only spatial location with a different behavior is above the penumbra, where the magnetic field drops slower than for the pores with their similar field strength at the photosphere, presumably because of the lateral expansion of the umbral fields. The curve of $dB/dz(z)$ for the penumbra cuts across some of the others in Figure \ref{fig17}.  
 
\begin{table}
\caption{Gradient of magnetic field strength $dB/dz(z)$ at $z=0$\,Mm and scale height of $B$.}
\hspace*{-1.5cm}\begin{tabular}{c|cc|cc}
region & original HMI & scaled HMI & \multicolumn{2}{c}{scale height [Mm]} \cr
 & G\,km$^{-1}$ & G\,km$^{-1}$ & NFFF  & POT  \cr\hline\hline
full FOV & -0.06 & -0.07 & 37 & 17 \cr
umbra & -0.50 & -0.58 & 8 & 8 \cr
penumbra & -0.25 & -0.35 & 11 & 10\cr
pores & -0.49 & -0.60 & 11 & 8 \cr
plage & -0.27 & -0.53 & 22 & 9 \cr
\end{tabular}
\label{tab1b}
\end{table}

\begin{figure}
\resizebox{7.5cm}{!}{\includegraphics{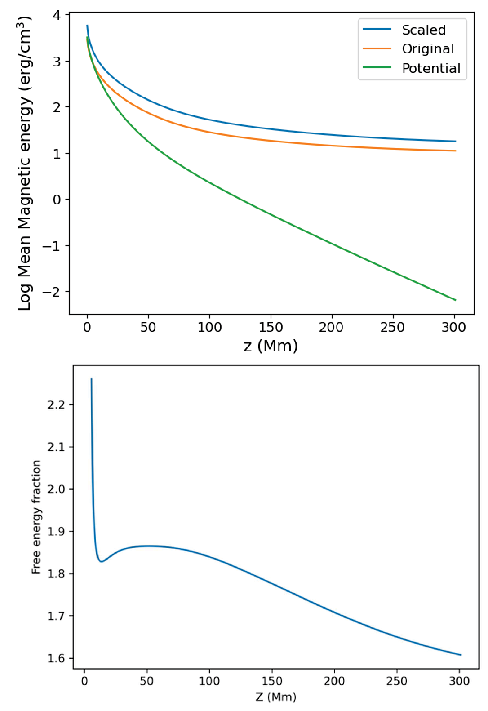}}
\caption{Mean magnetic energy (top panel) and ratio of the original and scaled free magnetic energy (bottom panel). }\label{fig16}
\end{figure}
\subsubsection{Free Energy}
The top panel of Figure \ref{fig16} shows the mean magnetic energy in the original and scaled NFFF extrapolation together with that of the potential field extrapolation. The mean magnetic energy in the latter turned out to be slightly larger than in the original NFFF extrapolation at low heights $< 10$\,Mm. We assume that this a consequence of the necessary pre-treatment of the HMI magnetogram to enforce a full magnetic flux balance for the potential field extrapolation that is not required for the NFFF extrapolations. It, however, does not impact the fact that the upscaled NFFF extrapolation has a higher mean magnetic energy than the original one. 

The bottom panel of Figure \ref{fig16} shows the ratio of the free energy between the upscaled and original NFFF extrapolation. The free energy was in each case derived by subtracting the magnetic energy in the potential field extrapolation $E=E_{NFFF}-E_{pot}$. The values were then horizontally averaged at each height $z$ prior to division, while additionally a minimum value of $10^{-10}$ was added to the averaged original free energy to prevent a possible division by 0. The values below about 10\,Mm are not reliable, as the free energy was negative. For the heights above, the upscaling leads to an increase by a factor of about 2 at $z=50$\,Mm that decreases to 1.6 at the upper boundary of the extrapolation box at $z=300$\,Mm, i.e., a 100 (60)\,\% increase, respectively.


\begin{figure*}
\resizebox{17.cm}{!}{\includegraphics{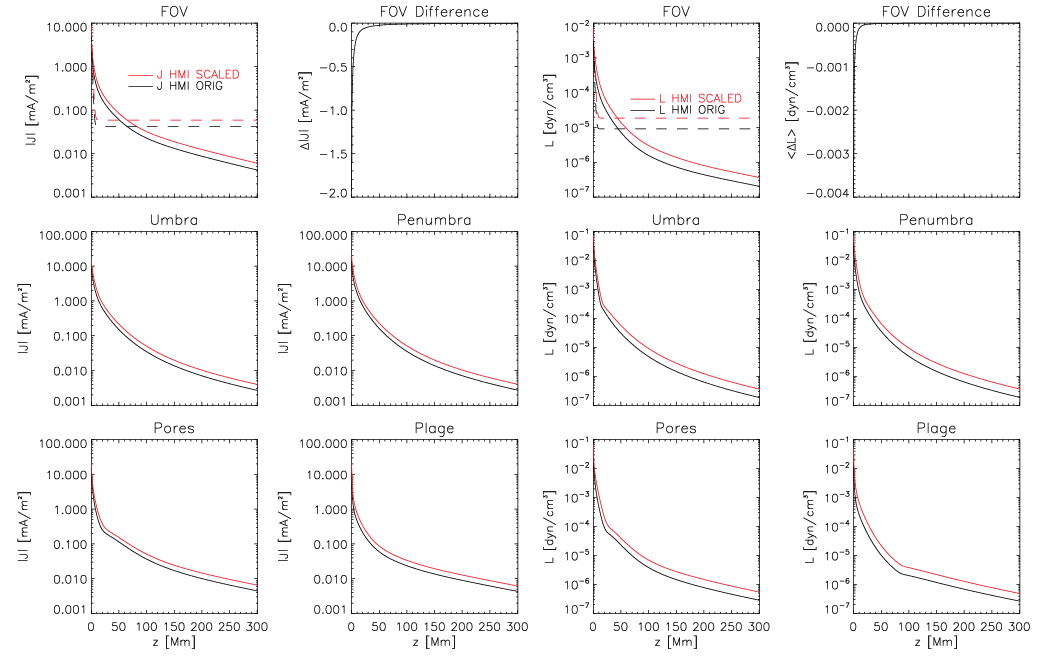}}
\caption{Average electric currents (left two columns) and Lorentz force (right two columns) in the same format as Fig.~\ref{fig15}.}\label{fig_curry}
\end{figure*}

\begin{table}
\caption{Electric currents $|J|$ and Lorentz force $L$ at $z=0$\,km.}
\hspace*{-2cm}\begin{tabular}{c|cc|cc}
region & $|J|$  & $|J|$ &  $|L|$   & $|L|$  \cr
 & mA\,m$^{-2}$ & mA\,m$^{-2}$ & $10^{-2}$\,dyn\,cm$^{-3}$ & $10^{-2}$\,dyn\,cm$^{-3}$\cr
 & original &scaled &original & scaled \cr\hline\hline
full FOV & 7 & 9 & 0.3 & 0.7 \cr
umbra & 12 & 14 & 5.0 & 7.3 \cr
penumbra & 15 & 21 & 3.7 & 7.4\cr
pores & 15 & 20& 3.5 & 6.4 \cr
plage & 11 & 21 & 1.5 & 4.9 \cr
\end{tabular}\label{tab_curry}
\end{table}
\subsubsection{Electric Currents and Lorentz Force}
The electric current vector $J$ is derived from the spatial gradients of the magnetic field strength as $J = \nabla \times B$, while the Lorentz force $L$ is given by $L = J \times B$. The modulus of $J$ and $L$ thus scales with the field strength itself. Figure \ref{fig_curry} shows the height variation of $J$ and $L$ in the full FOV and the different types of structures defined above. They both drop much faster than the field strength itself with a scale height of only about 1\,Mm (cf.~Equation \ref{exp_drop}) and a drop of about three orders of magnitude over the first 10\,Mm in height. The difference in $J$ between the original and scaled HMI field strengths has the same range as $B$ itself with a factor from 1.3 for the full FOV to 2 in plage (see Table \ref{tab_curry}) with the smallest changes seen above the umbra and the largest above plage regions. The Lorentz force shows a similar relation with respect to the different structures but its range of change is larger from 2--3 because of its dependence on both $J$ and $B$.     

\begin{figure}
\resizebox{8.cm}{!}{\includegraphics{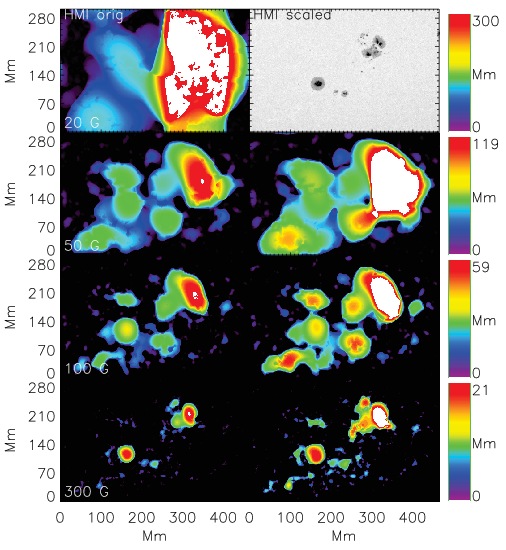}}
\caption{Height of different magnetic field strength values. Left column, bottom to top: height for $B=300, 100, 50$ and 20\,G in the original HMI data. Right column: the same for the scaled HMI data. The topmost panel has been replaced with the HMI continuum intensity as reference.}\label{fig18}
\end{figure}
\subsubsection{Height of $B = X$\,G}
A possible application of the combined use of magnetic field extrapolations and high-resolution observations in the chromospheric \ion{He}{i} line at 1083\,nm is to attribute a formation height to the \ion{He}{i} observations based on the field strength $B$ derived from them. The commonly used inversion codes for \ion{He}{i} at 1083\,nm only yield $B$ but usually provide no height estimate. The upscaling of the HMI magnetogram at the bottom boundary can then have a significant effect on the values of $B$ higher up, which would modify the estimate of the \ion{He}{i} 1083\,nm formation height.   

\begin{figure}
\resizebox{8.8cm}{!}{\includegraphics{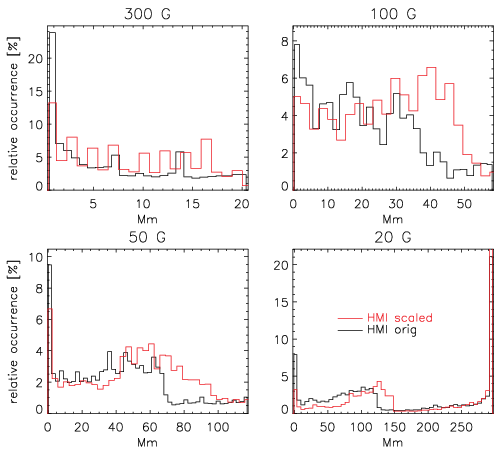}}
\caption{Histograms of the height of different magnetic field strength values. Clockwise, starting at lower right panel: for $B=20, 50, 300$ and $100$\,G. Black (red) lines: original (scaled) HMI data.}\label{fig19}
\end{figure}

Figure \ref{fig18} shows the heights in the extrapolation box where the field strength attains values of 20, 50, 100, and 300\,G as a common range of $B$ in \ion{He}{i} inversion results for the original HMI data (left column) and after the upscaling (right column). Field strengths of 300\,G or more can only be found above sunspots and below heights of about 20\,Mm. Values of 100\,G can be found up to about 1.5 the sunspot radius at heights of about 30\,Mm. Locations with 50\,G extend to about twice the sunspot radius at up to 60\,Mm, and can also be found above plage regions. 20\,G can be found throughout the full FOV up to the very top of the extrapolation box at 300\,Mm, but are primarily reached at about 150\,Mm. The histograms of the height where $B$ reaches these values in Figure \ref{fig19} show the common maximal height range more clearly, with clear drops of the occurrence rate at 20, 30, 60, and 150\,Mm for 300, 100, 50 and 20\,G, respectively. The histograms for the 20\,G-case (bottom right panel of Figure \ref{fig19}) show a maximum at $z=300$\,Mm because that value is reached over extended areas of the upper boundary of the extrapolation box (upper left panel of Figure \ref{fig18}) or on average for the upscaled HMI data (Table \ref{tab1a}). In addition, the matching height was determined as the pixel in $z$ that has the minimum difference in field strength to the specified value, which returns the upper end for all spatial positions that never drop below 20\,G at all heights. For the upscaled HMI data, the heights and the area in the FOV where they can be reached are increased. 

\begin{figure}
\resizebox{8.8cm}{!}{\includegraphics{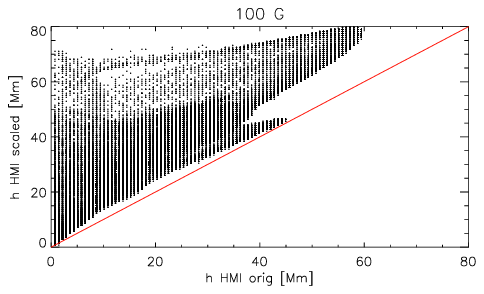}}
\caption{Scatter plot of the height with $B=100$\,G between orignal and scaled HMI data. The red line indicates unity slope.}\label{fig20}
\end{figure}

\begin{table}
\caption{Ratio of heights with B= X\,G between scaled and original HMI data.}
\begin{tabular}{c|cccc}
 & 300\,G & 100\,G& 50\,G & 20\,G \cr\hline\hline
h(scaled)/h(orig) & 3.16 & 2.45 & 2.99 & 4.92 \cr
\end{tabular}\label{tab3}
\end{table}

Figure \ref{fig20} shows a scatterplot of the heights with $B=100$\,G in the original and upscaled HMI extrapolations to determine the relative change with higher accuracy than from the 2D maps and histograms of Figures \ref{fig18} and \ref{fig19}. The height in the upscaled extrapolation increased in all cases by a factor 1--4. Table \ref{tab3} lists the average values of the ratio of the heights without and with upscaling for the four different field strength values. The height changes on average by a factor 2--5 with the upscaling of the HMI magnetogram at the bottom boundary. For the application of attributing formation heights to \ion{He}{i} 1083\,nm measurements based on a comparison of the field strength from an inversion and in an extrapolation the upscaling of the initial magnetogram thus seems clearly recommended. 

\section{Summary \label{sec_summ}}
From a comparison of HMI vector magnetic fields with those derived from observations with the high-resolution Hinode SP we find that the standard HMI ME inversion underestimates the field strength in all granular surroundings by a factor of 4--10. The primary reason for the mismatch are spatially unresolved magnetic fields that are not accounted for by a magnetic fill factor. The same effect is found within the SP data if a ME inversion with a unity fill factor is used. We derived a correction curve between HMI $B$ and the effective total flux $B\,f$ in the SP SIR inversion as HMI $B$ traces magnetic flux instead of field strength. The curve scales HMI $B$ up by a factor of about 2 for field strengths around 400\,G, dropping to unity at 220\,G and 2\,kG. All quantities in a large extrapolation box such as $B$, $J$ and $L$ scale correspondingly, while significant changes in the connectivity only happen for about 10\,\% of the field lines. There seem to be no obvious side effects of the scaling. The heights where $B$ reaches some given value increase by a factor 2--4.
\section{Discussion \label{sec_disc}}
\begin{figure}
\centerline{\resizebox{6.cm}{!}{\includegraphics{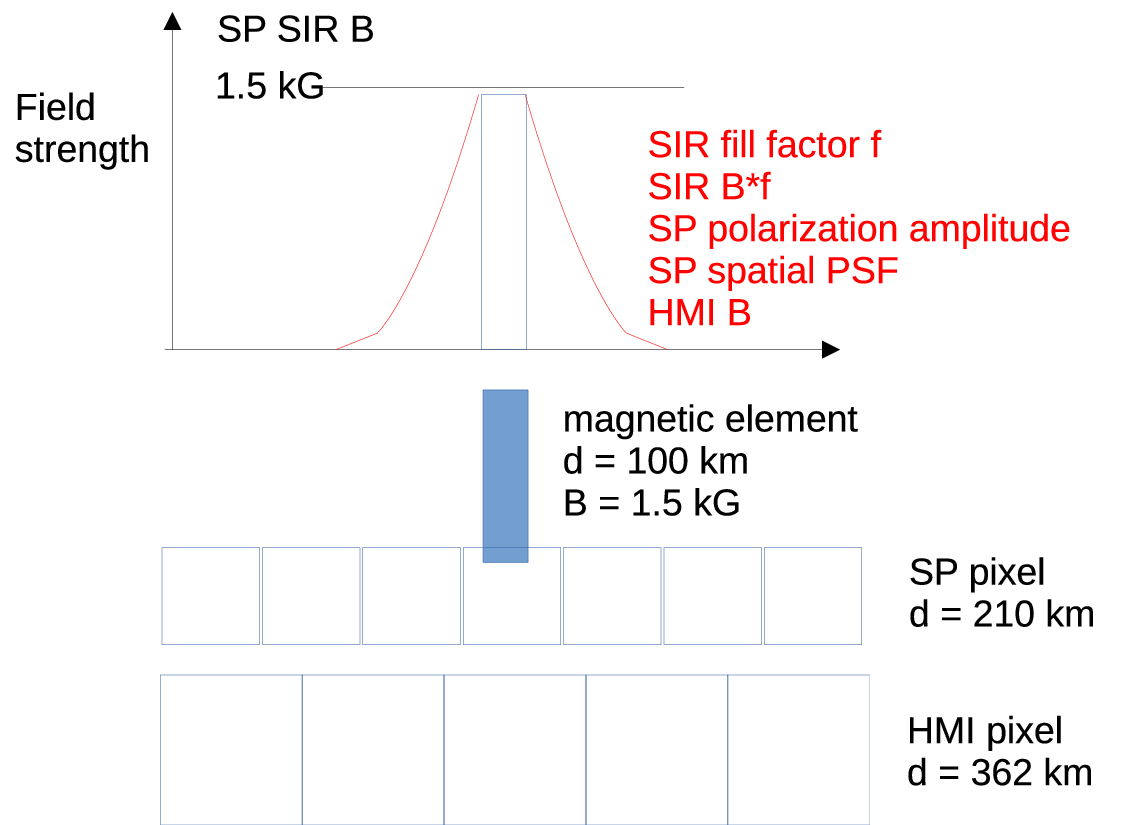}}}
\caption{Sketch of the geometry of an isolated magnetic plage element and its appearance in HMI and SP data. The spatial variation of all quantities mentioned in red apart from the SP field strength $B$ follows the shape of the red lines in the top part.}\label{fig9}
\end{figure}

\begin{figure}
\centerline{\resizebox{8.8cm}{!}{\includegraphics{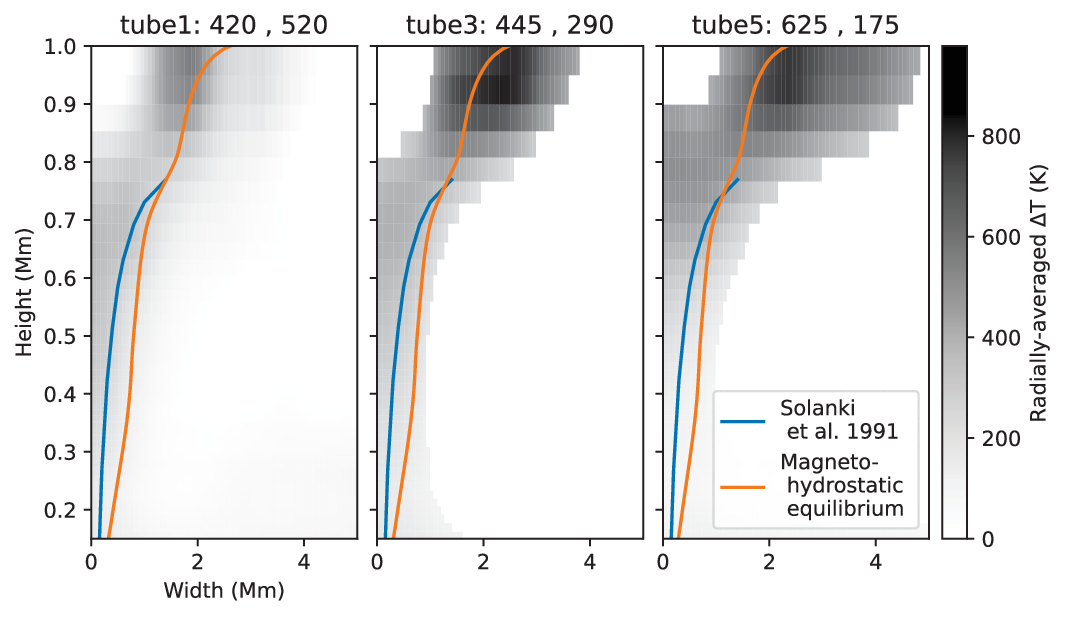}}}
\caption{Thermal canopy of magnetic elements in the QS. Left to right: three examples of the temperature around isolated magnetic elements in IBIS \ion{Ca}{ii} IR data. The blue solid line gives the outer boundary of the flux tube model of \citet{solanki+etal1991}. The orange line gives the boundary of a magnetic flux tube in magneto-hydrostatic equilibrium with the HSRA model. Courtesy of J. Jenkins.}\label{fig9a}
\end{figure}

\subsection{Behavior of Field Strength $B$ and Effective Flux $B\,f$}
The difference in the field strength $B$ in the HMI and SP data could be traced back to be mainly caused by the presence of unresolved magnetic fields in a pixel and its corresponding fill factor $f$. Figure \ref{fig9} visualizes the underlying effect. Isolated magnetic elements in the quiet Sun are usually located in the narrow intergranular lanes and have diameters down to below 100\,km \citep{berger+title2001,beck+etal2007,keys+etal2020}. They are unresolved in either the SP (0\farcs3\,pix$^{-1}$ for the current data) or HMI data (0\farcs5\,pix$^{-1}$) in a single pixel. In the absence of another source of polarized light, their polarization signal is spread out by the spatial point spread function \citep[e.g.,][]{staveland1970,martinezpillet1992,wedemeyer2008,mathew+etal2009,beck+etal2011} across their surroundings with a reduction of the polarization amplitude with increasing distance. Almost all quantities such as the polarization degree, the fill factor, the total flux or HMI $B$ have the same trend, apart from the field strength from the SP SIR inversion with a fill factor that stays at the central 1.5\,kG value. That behavior corresponds to the right column of Figure \ref{fig8} for a spatial cut through a plage region. The main consequence of the way HMI data are recorded and evaluated is that the reduction of the polarization amplitudes converts into a spurious reduction of field strength, while the SP SIR inversion can disentangle the ambiguity between field strength and fill factor for spatially unresolved fields because of using spectral lines at full spectral resolution.

Figure \ref{fig9a} shows that the spatial PSF is the primary reason for the behavior of the spread of polarization signal around magnetic elements, and not the magnetic canopy that forms in the chromosphere. The magnetic field spreads laterally in the upper atmosphere to maintain magneto-hydrostatic pressure equilibrium because of the exponential drop of the gas density \citep[e.g.,][]{solanki+etal1991,prasad+etal2022}. The lateral expansion in either the magnetic field \citep[taken from][their Model B with $B=1.6$\,kG and $d=100$\,km]{solanki+etal1991} or the thermal canopy \citep[from an NLTE inversion of Ca II IR spectra; see also][their Figure 11]{beck+etal2013c} becomes only significant at heights above $\sim 700$\,km. The formation heights of the spectral lines employed by HMI or SP are, however, limited to below 300\,km \citep{cabrera+bellot+iniesta2005,grec+etal2010,fleck+etal2011}, which is not high enough to sample the canopy. The magnetic canopy can thus not produce the lateral spreading or "blooming" of the field strength in the SP data.

Another strong indicator that HMI underestimates the field strength not only in the surroundings of but also in magnetic elements in a granular environment are the histograms of field strength in plage in Figures \ref{fig3} and \ref{fig_stray_corr} \citep[see also][his Figure 3]{sainzdalda2017}. The equipartition field strength, where kinetic and magnetic pressure are equal, is about 0.4\,kG in the photosphere in the quiet Sun. Network elements are stable for days to weeks and thus must have a field strength above that limit as found for the SP inversion with $0.5-1.8$\,kG for plage, while most of the HMI values are $<0.5$\,kG.

In total, we find that the HMI ME inversion results underestimate all field strength values $B$ in a QS environment by a factor of 3-10 because of the intrinsic way of acquiring and evaluating the HMI data, where small magnetic elements and their polarization signal are spatially {\it and} spectrally unresolved. 

The HMI LOS magnetic flux has the same scaling to the SP results as the magnetic field strength when the HMI ME full-vector inversion results are used. The inclination values $\gamma$ of both HMI and SP in the FOV (not shown) were very similar. The difference between $\Phi_{HMI}=B \cos \gamma A$ and $\Phi_{SP} = B f \cos \gamma A$ is thus the same as the scaling for $B$ and $B\,f$ in Figure \ref{fig10}. HMI LOS flux values derived at the 45\,s cadence from the simpler Stokes $I$ and $V$ measurement \citep{schou+etal2012} might be closer to the true magnetic flux value, but are prone to suffer in the same way from the lack of the fill factor for unresolved structures in the analysis. The basic assumption of the magnetograph equation in the weak-field limit that the amplitude of Stokes $V/I$ $\propto c\,B\,dI/d\lambda$ breaks down for unresolved fields, because for $f<<0.5$ the majority of the unpolarized photons $I$, and also $dI/d\lambda$, are not related to the source of the polarized photons anymore. The spectra with similar polarization amplitudes in Figure \ref{fig0} have different combinations of $B$ and $f$, but neither the product $B\,f$ nor their LOS magnetic flux $B f \cos \gamma$ are the same. At small $f$, the amplitude of Stokes $V/I$ strongly depends on the thermal stratification of the part of the pixel that does not host magnetic fields. Just by varying the temperature stratification, one can scale polarization amplitudes over a comparably large range of almost one order of magnitude at the same field strength and magnetic flux values \citep[][Appendix B]{beck+rezaei2009}.      
\subsection{Spatial Resolution and Fill Factor}
Using data of higher spatial resolution than the SP, improving its spatial resolution by a deconvolution with the PSF prior to the inversion \citep{beck+etal2011a} or using a spatially-coupled inversion scheme \citep{vannoort2012} would not resolve the discrepancy to HMI, but worsen it. A deconvolution of HMI data improves the situation \citep[Figure \ref{fig_stray_corr};][]{diazbaso+asensio2018}, but is finally still limited by the HMI pixel size of about 360\,km that even without any effects from the PSF cannot resolve magnetic elements with sizes of 100\,km, which would only have a fill factor of $f\sim 0.1$ in HMI data.

The primary way to achieve a closer match of HMI with the true values of field strength or magnetic flux would be to include a fill factor in the analysis of HMI data or simultaneously use data of higher spatial resolution to better constrain the results \citep{higgins+etal2022}. \citet{grinon+etal2021} demonstrated that the spectrally coarsely sampled HMI data still have enough independent information to use a fill factor in their analysis. They found that for weak magnetic fields outside of sunspots a modeling that includes a magnetic fill factor is strongly preferred, analogously to the results of $f$ for the SP data in Table \ref{tab1}. A possible other source of full-disk photospheric magnetic field information at a comparable spatial resolution would be the Synoptic Optical Long-term Investigations of the Sun \citep{keller+etal2003} as its 630\,nm slit spectra allow one to use a fill factor like for the SP data in the inversion of full-disk observations.
\subsection{Formation Height and Magnetic Field Gradients}
The photospheric \ion{Fe}{i} lines at 617.3\,nm and 630\,nm used by HMI and SP have a comparable formation height of $0-300$\,km \citep{grec+etal2010,fleck+etal2011}. Any variation of the optical depth scale across the FOV by, e.g., the Wilson depression in sunspots, will thus be similar and cannot explain the difference between HMI and SP. 

The spectral resolution of the SP data would allow us to use additionally magnetic field and velocity gradients in the inversion that are necessary for locations with significant net circular polarization \citep[e.g.][]{auer+heasley1978,almeida+lites1992,beck2011}. Given the typical values of the field strength gradients of $0.5-1$\,G\,km$^{-1}$ \citep[Table \ref{tab1b};][]{balthasar2018}, any inclusion of gradients would not change the result of the current comparison to HMI, since, e.g., a constant value of 1\,kG would only change to 850--1150\,G over 300\,km, which is insufficient to explain the discrepancies between HMI and SP.          
\subsection{Changes in the Magnetic Field Extrapolation}
The upscaling of the field strength in the lower boundary layer has a minor impact on the connectivity of the field lines in the extrapolation. Most ($\sim 90$\,\%) of them change in height or laterally by less than 10\,Mm, which implies, e.g., sunspots still connect to the same plage area, just at a slightly different location. Field strength $B$ and magnetic flux increase by a factor of about 2, while all dependent quantities such as electric currents, Lorentz force and free magnetic energy just scale accordingly. The main consequence is that any ARs with a large area fraction of plage instead of sunspots are predicted to be more strongly affected, or the other way around, HMI will underestimate both $B$ and $\Phi$ more in that case, especially since our scaling to the total effective flux $B\,f$ still falls short of the true field strength $B$. For our intented purpose of attributing formation heights to field strength values from an inversion of \ion{He}{i} 1083\,nm data through a comparison to a field extrapolation the scaling turns out to be necessary, as the corresponding heights more than double.
\subsection{Applicability and Limitations of Scaling Curve \label{lim_scaling}}
The current scaling curve between the field strength HMI $B$ and the total magnetic flux $B\,f$ in the SP data was derived using a single SP data set at a heliocentric angle of about 16 degrees. The FOV samples a broad variety of structures including fully formed sunspots, pores, plage and QS regions. We consider the result to be robust as far as different types of solar surface structures are concerned. The scaling uses only the initial value of HMI $B$ as the input to determine the scaling modulus and can thus in principle be applied to any HMI observations across the full solar disk. 

One caveat is that a possible dependence of the scaling on the heliocentric angle should best be tested with a similar data set at a preferably large heliocentric angle $>50$ degrees. A second limitation was found at the upper end of the scaling curve for $B>2400$\,G where the current value seems to be slightly too high for strong umbral fields and presumably should be again about unity instead. However, we consider the main limitation to be the artificial cut-off at low field strength values $B<220$\,G with a scaling coefficient of unity, while the actual results on average (Table \ref{tab2}) would suggest rather a larger value of $3-10$ at small HMI $B$ values. 

Without a quantity such as the polarization degree in the SP data that allows one to spatially filter out locations with genuine polarization signal opposite to random noise in the HMI data, the correction at the low end of the field strength range cannot be better determined. Using a scatterplot of the ratio of HMI $B$ and SP SIR $B\,f$ instead actually gave a scaling curve that basically exploded towards zero $B$, which could not be used. To include the low end of the field strength range in HMI $B$ would require the same filtering for genuine signals in HMI not only in the derivation but also the application of the scaling curve, where neither the HMI $B$ nor the HMI magnetic flux work well to define the filter since obviously genuine values -- all coherent spatial patches over a few pixels in the HMI data -- can have the same modulus in $B$ as the single-pixel noise pattern. 

Finally, the scaling to the total flux $B\,f$ is only somewhat of an intermediate crutch, but without a magnetic fill factor in both the HMI inversion results and the magnetic field extrapolation it is the best possible compromise. The main consequence is that the HMI $B$ values {\it even after} the upscaling still fall short of the true value of $B$.   
\subsection{The (Missing) Open Flux Problem}
Recent results on the magnetic field strength and flux from in-situ measurements by the Parker Solar Probe and prior missions or other derivations in the interplanetary space \citep[see][and references therein]{wang+etal2022,arge+etal2024} usually exceed the corresponding values at those locations based on magnetic field extrapolations of photospheric measurements by a factor of $2-4$. Our current results suggest that this could easily result from too low values of $B$ or $\Phi$ in granular surroundings in {\it all} magnetograms that do not use a fill factor in the data analysis because of the limited spatial resolution or spectral sampling of the corresponding instruments. 

To resolve the ambiguity between fill factor and field strength requires to spectrally resolve the thermal broadening to reliably determine the amplitude, shape and wavelength separation of Zeeman polarization components, but a single spectral line is sufficient for the purpose \citep{deltoroiniesta+etal2010}. While for locations with a small magnetic fill factor $f$ the commonly used weak-field approximation can break down and the polarization amplitude can decouple from both $B$ and $\Phi$ (Figure \ref{fig0}), the use of the Zeeman splitting for the determination of $B$ for low resolution data \citep{petrie2022} is also no solid solution. For locations with a field strength below $\sim 1.5$\,kG, the splitting of visible lines such as \ion{Fe}{i} at 630.25\,nm with a Land{\'e} coefficient of 2.5 is not yet proportional to the field strength \citep[][their Figure 7]{beck+etal2007} and the field strength values derived from the splitting always yield values above 1\,kG as the minimum \citep[\ion{Fe}{i} at 617.3\,nm;][their Section 4.2.3]{blancorodriguez+kneer2010}. Neither the weak-field approximation nor the strong field regime suffice to determine correct field strength or magnetic flux values for visible lines. Both assumptions are mutually exclusive, but hold at different locations in the QS, so any analysis approach based on solely either of the two approaches is strongly biased. Only an inversion of spectrally resolved data with a magnetic fill factor can reliably break up the ambiguity for spatially unresolved magnetic fields.      

Given that the majority of the solar surface is always covered by quiet Sun and network regions opposite to the $> 2$\,kG fields in sunspots, a better match of extrapolated and in-situ measurements of the interplanetary magnetic field will presumably only be achievable by improving the accuracy of the photospheric boundary values in the extrapolations. The two options would be an increase in spatial resolution to about 100\,km to ensure that even small magnetic elements are fully resolved ($f\equiv 1$) or to increase the spectral resolution so that a fill factor can be used in the derivation of the magnetic field strength in the initial photospheric magnetogram. In an extrapolation, the fill factor could possibly be implemented by an adaptive mesh for the pixel size near the bottom photospheric boundary, while at a height of about 1\,Mm a common grid size could again be used because of the lateral spread of the magnetic flux (Figure \ref{fig9a}).   
\section{Conclusions  \label{sec_conc}}
We find that the field strength $B$ in standard HMI ME inversion results underestimates the true field strength in all granular convective surroundings (quiet Sun, plage) by a factor of 4--10 wherever there are spatially unresolved magnetic field because of the lack of a fill factor in the inversion. The mean or total magnetic flux is underestimated by at least a factor of 2. Our scaling curve to match HMI $B$ and Hinode SP $B\,f$ has no obvious side effects on subsequent magnetic field extrapolations apart from a corresponding upscaling of $B$ and all quantities derived from it. The correction is based solely on the initial value of HMI $B$ and could thus be applied to any HMI data set without requiring simultaneous high-resolution observations for the application.    

Acknowledgements:\\
NSO is operated by the Association of Universities for Research 
in Astronomy (AURA), Inc. under cooperative agreement with the National Science Foundation (NSF). HMI data are courtesy of NASA/SDO and the HMI science team.  They are provided by the Joint Science Operations Center -- Science
Data Processing at Stanford University. {\it Hinode} is a Japanese mission developed and launched by ISAS/JAXA, collaborating with NAOJ as a domestic partner and NASA and STF (UK) as international partners. Scientific operation of the Hinode mission is conducted by the Hinode science team organized at ISAS/JAXA. This team mainly consists of scientists from institutes in the partner countries. Support for
the post-launch operation is provided by JAXA and NAOJ (Japan), STFC (UK), NASA, ESA, and NSC (Norway). We thank A.~Norton for providing us with the stray-light corrected HMI data. We thank J.~Jenkins for the graphics provided. D.P.C.~acknowledges support through the NSF grants AGS-1413686 and AGS-2050340. M.S.Y.~acknowledges support through the NSF grants AGS-2020703 and AGS-2230633. A.P.~would like to acknowledge support from the Research Council of Norway through its Centres of Excellence scheme (project number 262622) and Synergy Grant number 810218 459 (ERC-2018-SyG) of the European Research Council. Q.H.~acknowledges support through the NSF grant AST-2204385. We acknowledge using the visualisation software VAPOR (www.vapor.ucar.edu) for generating relevant graphics. We would like to thank the referee for helpful comments.

\bibliographystyle{aasjournal}
\bibliography{references_luis_mod}

\end{document}